\documentclass [psfig,12pt]{article}
\usepackage {psfig}

\def\refe{\par\noindent\hangindent=1.5cm}
\begin{document}
\def\basel{\normalsize\baselineskip=0.85truecm}
%\basel

\begin{center}

{\large \bf THE ANTHROPIC PRINCIPLE AND THE \\

\vspace{0.05cm}

DURATION OF THE COSMOLOGICAL PAST}

\vspace{1.5cm}

\large Milan M.\ \'Cirkovi\'c

\vspace{0.1cm}

{\it Astronomical Observatory Belgrade, \\
Volgina 7, 11160 Belgrade-74, Serbia and Montenegro \\

{\rm and} \\

Department of Physics, University of Novi Sad, \\ Trg Dositeja
Obradovi\'ca 4, 21000 Novi Sad, Serbia and Montenegro \\

E-mail: {\tt mcirkovic@aob.aob.bg.ac.yu}}

\vspace{1.7cm}
\end{center}

\begin{abstract}
The place of an anthropic argument in the discrimination between
various cosmological models is to be reconsidered following the
classic criticisms of Paul C. W. Davies and Frank J.\ Tipler.
Different versions of the anthropic argument against cosmologies
involving an infinite series of past events are analyzed and
applied to several instructive instances. This is not only of
historical significance but presents an important topic for the
future of cosmological research if some of the contemporary
inflationary models, particularly Linde's chaotic inflation, turn
out to be correct. Cognitive importance of the anthropic
principle(s) to the issue of extraterrestrial intelligent
observers is reconsidered in this light and several related
problems facing cosmologies with past temporal infinities are also
clearly defined. This issue is not only a clear example of the
epistemological significance of the anthropic principle, but also
has consequences for such diverse topics as SETI studies,
epistemological status of cosmological concepts, theory of
observation selection effects, and history of astronomy.
\end{abstract}

\vspace{0.7cm}

Keywords: history and philosophy of astronomy, cosmology: theory

\newpage

\textit{How many kingdoms know us not!}

\vspace{0.2cm} Blaise Pascal, "Thoughts" (207)

\vspace{1cm}

\section{Introduction}

The simplest division of all cosmologies is into two broad
classes: those postulating the eternal universe and those which
postulate some origin of the universe, or at least the part of it
that cosmologists are currently inhabiting. Eternal universes (and
here by eternal I mean those with no beginning or end, or even
only those with no beginning) are the only ones which could
pretend to adopt some sort of stationarity, a condition which is
of singular importance in many branches of physics (among other
issues because the law of energy conservation is closely connected
with a translational symmetry of time), and which is certainly
seen as greatly simplifying the solution of specific problems
everywhere. For a long period of time, after the dogma of Creation
in 4004 BC was abandoned, the universe has been considered
eternal, although great minds, such as Newton's, began to perceive
some of the difficulties associated with such a proposition (e.g.\
North 1965). The resistance to the opposing view (which eventually
became what is today called the standard cosmology) was
established during most of the nineteenth and early twentieth
century and is epitomized in the words of one of the pioneers of
modern astrophysics, Sir Arthur Eddington, who in his
authoritative Gifford lectures for 1927, published under the title
\textit{The Nature of the Physical World}, flatly stated: "As a
scientist, I simply do not believe that the universe began with a
bang."\footnote{Eddington (1928), p. 85. These words of Eddington
preceded for more than two decades the coining of the expression
``Big Bang'', so they should not be interpreted as a critique of a
particular model (after all, the first model which could, in a
loose sense, be called a Big Bang model, was constructed by
Lema\^{\i}tre only in 1931), but as rejection of the general
concept of originating of the world in a finite moment of time.}
From the end of the Middle Ages until the Hubble observational
revolution in the third decade of the twentieth century, the
stationary worldview has been in one way or another the dominant
one among the educated classes. This explains, among other issues,
the dramatic reaction of most of the scientific community,
including Lord Kelvin, Holmes, Eddington, Crookes, Jeans and
others, to the discoveries of Clausius, Boltzmann and others,
which all imply a unidirectional flow of time and physical change.
Interestingly enough, even during this epoch the idea, today one
of most investigated issues in physics, that the thermodynamical
arrow of time originates in cosmology, has occasionally surfaced
(Price 1996; \'Cirkovi\'c 2003a, and references therein).

The power of a stationary alternative to the evolutionary models
of the universe has been reiterated in particularly colorful form
during the great cosmological controversy in late 1940s, 1950s,
and early 1960s (Kragh 1996). Although during this period of
conflict between the Big Bang and the classical steady state
theories numerous and very heterogeneous arguments appeared on
both sides of the controversy, the argument based on the anthropic
principle was only explicitly formulated a decade after the
disagreements ended. As is well known, of course, the debate
ceased when empirical arguments persuaded by far the largest part
of the cosmological community that a universe of finite age is the
only acceptable theoretical concept. However, the argument based
on the anthropic principle has been further developed during the
1980s and has gained relevance in a new and developing field of
quantum cosmology (together with other aspects of anthropic
reasoning). The present paper is dedicated to detailed
consideration of the content and range of applications of that
argument. In spite of the huge volume of writings on the
philosophical aspects of stationary cosmologies (e.g.\ Hawkins
1971; Gr\"{u}nbaum 1991; Balashov 1994; Kragh 1996), this
particular argument has not been discussed in detail so far.

In modern physical cosmology, the position is reversed compared to
the situation in first decades of the XX century. The evolving
universe with definite beginning enjoys almost universal support,
at least in the last three decades (for a comprehensive reviews of
the field, see Weinberg 1972; Harrison 1973; Peebles 1993). Prior
to the discovery of the cosmic microwave background (henceforth
CMB) in 1965, at least one of the stationary theories---the
classical steady state model of Bondi and Gold (1948), as well as
Hoyle (1948)---has been a viable and quite popular worldview.
After the interpretation of CMB as the remnant of the primordial
fireball (Dicke et al.\ 1965) has become a standard one (Sunyaev
and Zeldovich 1980), the evolutionary paradigm became universally
dominant.

However, this is not the end of the story. There are at least two
reasons (apart from appreciable historical interest) to study
arguments pertaining to the cosmologies with infinite past even
after 1965, both dealing with the boundary conditions, but with
significantly different slant.
\begin{itemize}
\item Great pains have been taken to make the initial singularity
palatable or to avoid it whatsoever (e.g.\ Misner 1969; Bekenstein
and Meisels 1980; Israelit and Rosen 1989), and the results are
still inconclusive from the point of view of most researchers, in
spite of the tremendous advances of the new discipline of quantum
cosmology. The issue of a compulsory nature to the initial
singularity in the classical context has remained a problem ever
since the famous singularity theorems (e.g. Hawking and Penrose
1970). This has been aggravated in view of the very special nature
of such a singularity entropy-wise (Penrose 1979; \'Cirkovi\'c
2003a). The significance of such a discourse is emphasized by
recent attempts to build an atemporal worldview (Price 1996), in
which cosmological boundary conditions play the crucial role. It
is particularly interesting to consider in this context
counterfactual cosmological models (such as the classical steady
state), and to compare their temporal aspects and their boundary
conditions with the realistic ones.\footnote{For instance, if the
thermodynamical arrow of time depends on the cosmological boundary
conditions, as suggested several times since Boltzmann, the notion
of time in the everyday world ultimately depends on the
low-entropy nature of the initial Big Bang singularity. However,
are we to discard the existence of an arrow of time in a steady
state universe in which there is no initial singularity and no net
increase in entropy?}

\item Recently, various "multiverse" schemes have been proposed
(e.g.\ Linde 1990; Smolin 1992), in which our visible universe is
only part of a larger structure (for a partial list of such
theories, see Bostrom 2002). Motivations for this sweeping
generalization have been multi-fold, ranging from details of the
inflationary theory to topology to the fundaments of quantum
mechanics. Some of the proposed multiverses are indeed in a
stationary state, even somewhat resembling of the classical
steady-state theory. If these, still rather novel, propositions
are to be taken seriously, their impact on several distinct
cosmological concerns have to be investigated. One of the problems
pertaining to the question "one or many universes" is that of the
validity of various anthropic arguments, among them the argument
against an infinite series of past events as discussed here.
\end{itemize}

The modern version of the anthropic argument against the past
infinite series of events (or past temporal infinity in
relationist terms; see the discussion below) has appeared in a
short notice by Paul C. W. Davies appearing in \textit{Nature }in
June 1978 (Davies 1978). In this succinct critique of the Ellis et
al. (1978) static cosmological model he points out that
\begin{quotation}
there is also the curious problem of why, if the Universe is
infinitely old and life is concentrated in our particular corner
of the cosmos, it is not inhabited by technological communities of
unlimited age.
\end{quotation}
\noindent As mentioned by Barrow and Tipler (1986) in their
encyclopaedic monograph, this is historically the first instance
in which an anthropic argument has been used against a cosmology
containing a past temporal infinity, and it is indeed fascinating
that nobody had considered it before. The suprise is strengthened
by the fact that such cosmologies in scientific or half-scientific
form have existed since the very dawn of science. Simultaneously,
since ancient times a belief in the existence of other
\textbf{inhabited} worlds has been present, in one form or
another.\footnote{For a historical sketch from the pen of a
``contact pessimist'' see Tipler (1981).} Today, the scepticism
sometimes encountered against this mode of thinking is even
stranger, when various (and in some cases not quite inexpensive)
SETI projects testify to the reasonable degree of belief in the
existence of technological civilizations other than the human one.
Their technological nature (the same one which produces the
problem Davies wrote about) is a \textit{conditio sine qua non} of
any sensible SETI enterprise.

In tight connection with this issue is the definition of anthropic
principles. Although we shall later discuss some other anthropic principles,
for the moment it is enough to define the weak anthropic principle
(henceforth WAP), which states (Carter 1974) that

\begin{quotation}
\noindent ...we must be prepared to take account of the fact that
our location in the universe is \textit{necessarily} privileged to
the extent of being compatible with our existence as observers.
\end{quotation}
\noindent An alternative definition is given by Barrow and Tipler
(1986):
\begin{quotation}
\noindent The observed values of all physical and cosmological
quantities... take on values restricted by the requirement that
there exist sites where carbon-based life can evolve and by the
requirement that the Universe be old enough for it to have already
done so.
\end{quotation}
With these formulations of WAP in mind, it is clear why the Davies' argument
against the Ellis et al. cosmological model can be called \textbf{anthropic}
at all: it takes into account the restrictions to be imposed on cosmological
models following the existence of a specialized subclass of intelligent
observers, namely the ``technological communities''.

An ancient echo of this type of argumentation can be recognized in
the surviving fragments of some of the most distinguished ancient
philosophers of nature. From our point of view especially
interesting is the cyclic cosmology of Empedocles of Acragas (VI-V
century BC), in which the universe is eternal,\footnote{It seems
clear that Empedocles held a sort of the absolutist theory of the
nature of time. In particular, the fragment B16 of the Diels
collection reads (according to the translation of Burnet 1908):
``For of a truth they (Strife and Love) were aforetime and shall
be; nor ever, methinks, will boundless time be emptied of that
pair.''} consisting of the internally immutable four classic
elements, as well as two opposing forces (Love and Strife, i.e.\
attractive and repulsive interactions). The cyclic motion of
matter in the universe is governed by the change in relative
intensities of the two interactions (see the excellent discussion
in O'Brien 1969). It is interesting to note that Empedocles'
cosmology is \textbf{uniformitarian}, in the sense that all six
basic constituents (four elements and two forces between them) are
present in each instant of time in accordance with the eternal
principles of mutual exchange. In some of the surviving fragments,
Empedocles implies that, although this uniformitarianism may seem
counterintuitive, as we see things coming into being and
vanishing, this is just our special perspective (today we would
say \textbf{anthropocentrism} or observation \textbf{selection
effect}) and not the inherent state of nature.\footnote{Another
pioneering contribution of Empedocles lies exactly in separation
(the earliest one in the Western thought!) of the physical nature
and artifacts of human cognizance. See, for instance, the Diels'
fragment B8, reading (in Burnet's translation): ``There is no
coming into being of aught that perishes, nor any end for it in
baneful death; but only mingling and change of what has been
mingled. Coming into being is but a name given to these by men.''
Even more telling along the same lines are fragments B11 and B15.}
This is strikingly similar to the uniformitarian notions present
in some of the most authoritative cosmological models of the
twentieth century, and we shall return to it in subsequent
discussion of the classical steady state theory (Balashov 1994).

As discussed in some detail in \'Cirkovi\'c (2003b), this
Empedoclean view---that biological evolution and the appearance of
consciousness and intelligence are contingent upon cosmological
processes---coupled with the notion of the eternal universe leads
to the same sort of trouble as the one facing the classical steady
state theory or the one of Ellis et al. criticized by Davies. Why
then, in the supposed infinity of time, are ``men and women,
beasts and birds'' of finite, and relatively small, age? Where are
traces of previous infinite cycles of the "world-machine" (cf.\
the discussion of Hutton in \S \ref{hutt} below)?

Empedocles may have perceived this himself and he evades the
problem in the only natural way he can: by postulating two
singular states in the beginning and in the middle of each of his
great cycles. These singular states are moments (in the absolute
time!) of complete dominance of either Love (an ancient equivalent
of the modern initial and/or final singularities) or Strife (no
true equivalent, but similar to the modern version of heat death
in the ever-expanding cosmological models; see, for instance,
Davies 1994). In these states the life, with its complex
organizational structure, is impossible and therefore they serve
as \textit{termini} for the duration of any individual history of
life and intelligence. In other words, the information about
anything that was before is destroyed in the singular events.
Strikingly, {\it the maximal duration of any form of life and/or
intelligence is determined exclusively by cosmological laws!\/}
Therefore, there are no arbitrarily old beings, and anthropic
argument is inapplicable. We shall meet the same strategy over and
over again in the history of cosmological ideas.

In the very first chapter of the history of Thucydides, there is a
famous statement that before his time---i.e. about 450
BC---nothing of importance ($\sigma \upsilon  \quad \mu
\varepsilon \gamma \alpha \lambda \alpha  \quad \gamma \varepsilon
\upsilon \varepsilon \sigma \theta \alpha \iota$) had happened in
history. This startling statement has been called ``outrageous''
by Oswald Spengler, and used to demonstrate the essentially
mythological character of the ancient Greek historiography
(Spengler 1918; see also Cornford 1965). It may indeed be
outrageous from the modern perspective, but it does motivate a set
of deeper questions, ultimately dealing with cosmology. The fact
that Thucydides did not know (or did not care to know) about
previous historical events does not change the essential
perception of \textbf{finiteness} of human history inseparable
from the Greek thought. This property starkly conflicts with the
notion of an \textbf{eternal continuously existent world}, as it
was presented in both modern and ancient cultures. Obviously, it
is irrelevant which exact starting point we choose for unfolding
historical events. In any case, the number of these events is
finite, and the timespan considered small even compared to the
specific astronomical timescales (some of which, like the
precession period of equinoxes, were known in the classical
antiquity, as is clear from the discussion in \textit{Timaeus}),
not to mention anything about a past temporal infinity. Although
there was no scientific archaeology in the ancient world, it was
as natural then as it is now to expect hypothetical previous
civilizations inhabiting Oikumene to leave some traces---in fact,
an infinite number of traces for an eternally existent Oikumene!
There are indications that pre-Socratic thinkers have been aware
of the incompatibility of this ``Thucydidean'' finiteness of
historical past with the eternal nature of the world. We have
already mentioned the solution (periodic singular states) proposed
by Empedocles himself. Even earlier, in the fragmentary accounts
of the cosmology of Anaximandros, we find an evolutionary origin
of humankind in some finite moment in the past, parallel with his
basic postulate of separation of different worlds from
\textit{apeiron} and their subsequent returning to
it.\footnote{This is clear, for instance, from the fragment A 10
in Diels (1983), preserved by Plutarch, in which it is explicitly
asserted that formation and destruction of many worlds occurs
within the global temporal infinity. In the continuation of the
very same excerpt from \textit{Stromateis}, an evolutionary
doctrine is attributed to Anaximandros: ``...Farther he says that
at the beginning man was generated from all sorts of animals,
since all the rest can quickly get food for themselves, but man
alone requires careful feeding for a long time; such a being at
the beginning could not have preserved his existence.'' (Fairbanks
1898) Hyppolites quotes Anaximandros as emphasizing the nature of
\textit{apeiron} as eternal (B 2), obviously in opposition to
mankind, which has a fixed beginning in time. Even more intriguing
is the doctrine ascribed to Anaximandros by Cicero: ``It was the
opinion of Anaximandros that gods have a beginning, at long
intervals rising and setting, and that they are the innumerable
worlds. But who of us can think of god except as immortal?'' Did
he have in mind essentially what we today denote as
supercivilizations?} In Anaxagoras' worldview, there is a famous
tension between the eternity of the world's constituents and the
finite duration of \textbf{movement} (and, therefore, relational
time) in the world. In the same time, it seems certain that
Anaxagoras, together with Anaximandros and Empedocles, was an
early proponent of the evolutionary view, at least regarding the
origin of humankind (Guthrie 1969).

Subsequently, with the Epicurean school, this issue became an
argument for the finite origin of the universe. Most eloquently,
it has been put forward in Roman times by Lucretius, who in the
Book V of his famous poem \textit{De Rerum Natura} writes the
following intriguing verses:

\begin{quotation}
Besides all this,

If there had been no origin-in-birth

Of lands and sky, and they had ever been

The everlasting, why, ere Theban war

And obsequies of Troy, have other bards

Not also chanted other high affairs?

Whither have sunk so oft so many deeds

Of heroes? Why do those deeds live no more,

Ingrafted in eternal monuments

Of glory? Verily, I guess, because

The Sun is new, and of a recent date

The nature of our universe, and had

Not long ago its own exordium.\footnote{In translation of William
E. Leonard, available via WWW Project Gutenberg (Lucretius 1997).}
\end{quotation}

\noindent For scientific-minded Lucretius, the shortness of human
history \textbf{is} very strange on the face of hypothesis of the
eternal existence of the world. Although the reference to
``eternal monuments'' may sound naive, it is clear that he had in
mind any form of transmission of information from the past to the
present; and an infinite amount of information from an infinite
past. His empirical assessment clearly shows the absence of such
information. Therefore, an explanation is needed. The simplest
explanation, as Lucretius was highly aware, is to assume that the
world is of finite (and relatively small) age. This is
\textbf{exactly} what modern cosmologists Davies and Tipler have
had in mind when constructing the anthropic argument.

\section{Davies-Tipler argument: modern formulation}

The anthropic argument against steady-state theories hinted at by
Davies in the quotation above has been subsequently expanded and
elaborated by Tipler (1982), and it seems only just to refer to
its modern form as Davies-Tipler (henceforth DT) argument. In the
latter work it has been shown that this argument applies to "all
universes which do not change with time in the large", and
particularly those which satisfy the so-called perfect
cosmological principle (Bondi and Gold 1948; see the discussion in
the Sec. 3 below). The discrete Markov chain recurrence of the
type discussed by Ellis and Brundrit (1979) has also been used in
the discussion of Tipler (1982), although, as we shall discuss
below, its use is largely superfluous, since even a much weaker
hypothesis produces the same disastrous effects for the
cosmologies with past temporal infinities.

The essence of Tipler's (1982) discussion is that, given some
usual symmetries of spacetime, for each event $p$, its past light
cone intersects all world lines corresponding to the history of an
intelligent species. Thus, at least one out of an infinite number
of such species, could travel along the time-like geodesic to $p$
(or just send signals). Since $p$ may be any event, like our
reading of Tipler (1982) paper, or any other occurrence in the
Solar system, it is completely unexpected that we are not already
part of an intelligent community of an arbitrarily long age. For
the case of the universe satisfying the perfect cosmological
principle (i.e. the classical steady state universe) it is clearly
seen from the Penrose diagram shown in Fig. 1. Again, it is
important to stress \textbf{non-exclusivity} of this argument:
even if 99.99{\%} (or indeed any fraction less than unity) of
intelligent communities arising at, say, $q$ would not expand
further than some limited neighborhood \textit{q + $\varepsilon
$}, in an infinitely old universe there would still be at least
one intelligent community at any point $p$ in spacetime, no matter
how big \textit{$\vert $p - q$\vert $ / $\varepsilon $} is.

Here we notice that instead of dealing with temporal parts, we are
dealing with events in spacetime only. Therefore, this argument is
intrinsically formulated in terms of a relationist or reductionist
theory of time (e.g. Newton-Smith 1980). However, it is not
contingent upon acceptance of any particular theory of time. Both
relationist and absolutist pictures can accommodate the anthropic
argument, provided that some additional specific requirements are
satisfied. Conversely, if these requirements are not satisfied,
the argument is inapplicable, no matter whether we regard time as
contingent upon world events or an absolute background to any
event. In the absolutist picture, we need to speak of infinite
number of non-trivial past events \textbf{in} time, instead of the
time itself, which remains completely irrelevant to the argument.
If such a series of events satisfy some additional constraints
(like requirement to include the creation of life and intelligence
among the physically possible events, etc.), we are led to the
same paradoxical conclusion.

\begin{figure}
\begin{center}
\psfig{file=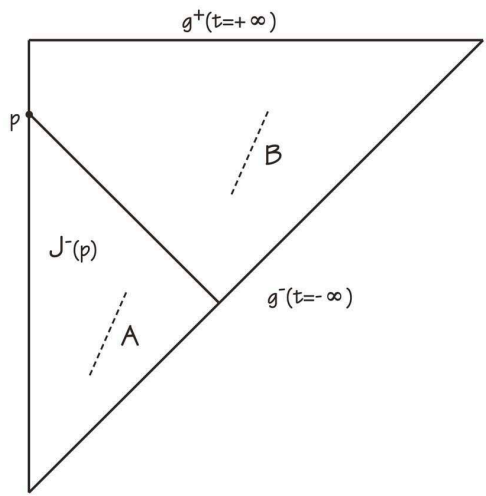,width=10cm}%,angle=270}
\vspace{-3cm}
\end{center}
\caption{Conformal (Penrose) diagram of the
classical steady-state universe. Event $p$ is arbitrarily chosen,
and with A and B are denoted worldlines of intelligent species
existing within and outside of the light cone of $p$. (Reproduced
with permission from Tipler 1982.)} \label{cetri}
\end{figure}

However, Tipler (1982) goes farther than Davies' casual remark
given in the context of an editorial comment, and claims that,
\begin{quotation}
\noindent [s]ince all possible evolutionary sequences have
occurred to the past of $p$, one of these evolutionary sequences
consists of the random assembly, without assistance of any
intelligent species whatsoever, of a von Neumann probe out of the
atoms of interstellar space. Such a random assembly would occur an
infinite number of times to the past of $p$, by homogeneity and
stationarity in an infinite universe. At least one of these
randomly assembled probes would have the motivations of a living
being, that is to expand and reproduce without limit.
\end{quotation}

This scenario, although not at all fantastic, raises several
questions still lacking elaboration. How could we possibly know
that the set of all "favorably-motivated" spontaneously assembled
von Neumann probes is of non-zero measure in the set of all such
probes. The question of motivation, which is not so easily
quantifiable, becomes crucial here. For instance, why not
postulate an assembly of von Neumann probe designed to search and
destroy other von Neumann probes? What is the relative weight of
colonizing (vs. destructive, altruistic, introvert, etc.)
motivation, and how can one determine it? This motivation problem
is avoided if we stick to a more restrictive requirement that only
communities of evolved intelligent beings create such probes (i.e.
create them at timescales many orders of magnitude shorter than
those required for the spontaneous assembly Tipler describes).
While one may argue that motivation is necessarily linked to the
level of complexity, and therefore one expects the spontaneously
assembled self-reproducing automata will have basically the same
motivations we perceive in biological systems on Earth (Tipler
2001, private communication), this issue is not clear at all.

Among the precursors of the anthropic argument of Davies and
Tipler, one may list the great British biologist, chemist,
philosopher and author John B. S. Haldane (1892$-$1964). His keen
interest in cosmological issues has been characterized by his
defense of the Milne's cosmological model in which (at least
according to one timescale) the universe is of finite age, and
fundamental constants change with time. In the following
interesting passage, through comparing the hypothesis of the
origin of the universe in finite past \textit{vs.} the hypothesis
of its eternal existence, he shows both his cosmological interests
and appreciation for a melioristic and humanistic worldview:

\begin{quotation}
\noindent On the first hypothesis, why was it not created better;
on the second, \underline {why has it not got better in the course
of eternity}?... On neither theory have we very strong grounds for
hoping that the world will be a better place a million, let alone
a thousand, years hence, than it is today. But on Milne's theory
the laws of nature change with time. The universe has a real
history, not a series of cycles of evolution. Although, from one
point of view, the past is infinite, life could not have started
much before it did, or have got much further than it has at the
present date. If this is so, human effort is worth while and human
life has a meaning.\footnote{Haldane 1945; underlined by the
present author.}
\end{quotation}
\noindent If we understand ``improvement'' of the universe not in
strictly ethical terms,\footnote{But presumably including that
aspect. If we reject pantheism, there is a minimal level of
complexity necessary for the subject and very notion of ethics to
exist. Therefore, any ethically melioristic cosmos must satisfy
specific WAP constraints.} but as increase in its
\textbf{complexity}, the question posed by Haldane is the same as
in the DT argument. Complexity may be achieved through either
technologization or ``biologization'' of the universe, and both
lead to paradoxical consequences.

\section{Two versions and related arguments}

Let us introduce the following terminology. By the \textbf{weak
DT} \textbf{argument}we shall denote the version introduced by
Davies in his 1978. note:

\vspace{0.3cm}

\noindent {\bf Weak DT:} \textit{Cosmologies postulating
inhabitable past temporal infinities must be rejected due to the
absence of traces of arbitrarily old civilizations in our past
light cone. }

\vspace{0.3cm}

The \textbf{strong version} is the one presented in Tipler (1982),
as well as in monograph of Barrow and Tipler (1986). Let us
formulate it by analogy with the weak argument in the following
form:

\vspace{0.3cm}

\noindent {\bf Strong DT:} \textit{Cosmologies postulating past
temporal infinities must be rejected for absence of activities of
spontaneously assembled self-reproducing automata in our past
light cone, provided that matter in our past light cone satisfies
constraints enabling such spontaneous assembly. }

\vspace{0.3cm}

In spite of the clumsiness of the latter formulation, it is clear,
upon careful inspection, that the auxiliary assumptions in the
stronger version are significantly more general than those in the
weak version. As far as both versions are concerned, it is very
important to notice that they do not guarantee the existence of
the entities under consideration. Let us consider the weak version
first as the more important from the overall point of view of this
work. It contains two stringent requirements which have been
usually tacitly assumed (and played an important historical role),
but which should be explicitly discussed:

\begin{enumerate}
\item \textbf{The requirement of continuous inhabitability of at
least a large enough region of causally connected space.} This
should be regarded as a form of restriction on spatial
inhomogeneities of the cosmological model under consideration. The
question what counts as inhabitable is actually a very subtle one,
since requirements of communities of intelligent observers may
wildly differ. In any case, we expect that a sufficiently large
thermodynamical disequilibrium must exist. Therefore, the question
whether a particular model satisfies this auxiliary assumption is
contingent upon the answer from the same model to the much more
famous problem of Olbers' paradox. We shall see below how various
specific cosmological models enable continuous entropy production
over an infinite interval of time.

\item \textbf{The existence of arbitrarily old civilizations.} The
existence of such entities can hardly be considered obvious or
even probable on any count. While the condition \ref{eq1}. is
necessary, it is not a sufficient reason for accepting \ref{eq2}.
Obviously, apart from the physical environment, there are other
reasons of a subtler and less quantifiable nature which could put
an upper limit to the age and growth of civilizations even in a
continuously inhabitable universe.
\end{enumerate}

Now we perceive what is the ultimate recourse to the proponents of
large-scale stationarity (as well as opponents of the usage of
anthropic principle for discriminating between the cosmological
models). While the rejection of either \ref{eq1}. or \ref{eq2}.
may seem easy enough, the careful inspection shows that it is a
rather difficult endeavor. Models which are characterized as
stationary automatically satisfy \ref{eq1}., and the reasons
sometimes cited for rejection of \ref{eq2}.---mostly in the
context of the SETI debate (see below)---are entirely \textit{ad
hoc}. Non-stationary models may violate \ref{eq1}., while
retaining infinite number of past temporal events, although this
violation is necessarily limited by spatial and temporal scales
our observations are probing. Limits following from the
observations are already too relaxed for us to conclude that this
is a sufficient reason for the rejection of \ref{eq1}., even apart
from the fact that no infinitely old non-stationary model has ever
been investigated in much detail.

The strong DT argument necessarily endorses a following auxiliary
assumption, analogous with the \ref{eq1}. in the weak case:

\begin{enumerate}
\item ' \textbf{The requirement of a continuously existing region
in which the spontaneous assembly of matter can be achieved.} This
presumes (i) availability of matter (so that empty models, like
the de Sitter world, can be excluded), and (ii) restrictions on
the physical state of matter. This second restrictions is hard to
make precise in the general case, but it seems clear that we
should exclude cases, for instance, in which the matter is at
temperatures high enough for any level of organization to be
immediately destroyed by thermal motions.
\end{enumerate}

It is generally much more difficult to construct an infinitely old
cosmological model violating \ref{eq1}' so that it can be brought
even into a very superficial accord with the empirical data which
any realistic theory must satisfy. However, since the physical
issues involving a spontaneous assembly of the desired sort are
not clear, we shall in further discussion concentrate on the weak
version of the argument, while only occasionally referring to the
stronger version. More stringent requirements \ref{eq1}. and
\ref{eq2}. produce make, for that matter, any hunt for loopholes
in the argument itself (and our consequent increase in
understanding) much more promising.

For the sake of better understanding of the issues involved, let
us consider a counterexample of a cosmological model involving
past temporal infinity which DT argument does not apply
to.\footnote{For a preliminary treatment of this topic, see
\'Cirkovi\'c 2000.} This is the Lema\^{\i}tre-Eddington universe,
which was quite popular in the 1925-1935 period (e.g. Eddington
1930). This model belongs to the class of general-relativistic
models with cosmological constant and without the Big Bang.
Therefore, it was very appealing from the point of view of
resolving the age discrepancies between cosmological models and
various astrophysical and geological timescales (Bok 1946; Kragh
1996). A good description of this model can be found in Bondi's
classic textbook on cosmology (Bondi 1961). Having appeared on the
cosmological scene after the realization of the instability of the
original Einstein static universe (Einstein 1917), this model

\begin{quotation}
\noindent ...has therefore an infinite past which was spent in the
Einstein state. This has greatly attracted investigators since it
seemingly permits an arbitrarily long timescale of evolution. The
picture of the history of the universe derived from this model,
then, was that for an infinite period in the distant past there
was a completely homogeneous distribution of matter in equilibrium
in the Einstein state until some event started off the expansion,
which has been going on at an increasing pace ever since. The
condensation of the galaxies and the stars from the primeval
matter took place at the time the expansion began, but this
development was stopped later by the decrease of average density
due to the progress of the expansion.\footnote{Bondi (1961), p.
118.}
\end{quotation}

This model is a good physical representation of the situation
often considered in philosophical studies of distinction between
the relationist and absolutist theories of time: the situation in
which an absolutely unchanged universe suddenly transforms into
changing world we observe (e.g.\ Hinckfuss 1975). From the formal
point of view, in accordance with the Weyl postulate, the
Eddington-Lema\^{\i}tre universe has an infinite past, i.e.\ the
initial state is given by the formal limit $t \to -\infty$.
However, this is a ``false'' infinity, at least in the context of
anthropic reasoning, because the period of time in which there are
conditions enabling the creation of intelligent observers is
necessarily finite. In addition, this period is approximately
equal to the time past since the beginning of the expansion. In
the Leibnitz-Berkeley-Machian relationist picture, the time itself
does not really exist before the onset of instabilities, i.e. the
universal expansion. The period of complete homogeneity can be
regarded as a state analogous to the epochs of complete dominance
of Love or Strife in the cosmology of Empedocles or, even more
accurately to the time before the motion began in Anaxagoras'
cosmology.\footnote{The fragment B64 of Diels suggests that
Anaxagoras endorsed a version of what were to become the
absolutist or Newtonian (or Platonist in Newton-Smith's parlance)
theory of time. In this description of Anaxagoras' cosmology,
preserved in Simplicius' commentaries on Aristotle's
\textit{Physics}, it is emphasized that the world itself was
created at the beginning of motion, but that {\bf before that
time} everything was static. As in Eddington-Lema\^{\i}tre
universe, the time axis can be formally extended backwards to $-
\infty $.} In both cosmologies it is necessary to invoke a state
which prevents propagation of information from an arbitrarily
distant past to the present epoch. In both cases this goal is
achieved by postulating states with a sufficiently high degree of
symmetry.\footnote{Of course, this interpretation of the
Eddington-Lema\^{\i}tre universe is not mandatory. It may as well
be said that stars and galaxies have existed indefinitely as such
before the expansions starts. Of course, in this case the
anthropic argument becomes valid at any epoch in the finite past,
even before the start of the expansion, since it reduces to the
application in the static Einstein universe. However, there are
additional arguments against such interpretation of the
Eddington-Lema\^{\i}tre model. For instance, as shown in an
instructive study by Pegg (1971), the model with indefinite past
containing stars can be rejected as a consequence of a very grave
form of the Olbers' paradox.} Obviously, in the case of the
Eddington-Lema\^{\i}tre universe, the anthropic argument is
inapplicable, since the \textbf{effective past} is finite.
Intelligent observers (or spontaneously assembled von Neumann
probes!) possess only a finite time for technologization of their
cosmic environment. This is valid for the generic version of the
Eddington-Lema\^{\i}tre model. Of course, the model pretending to
describe the real universe is normalized to the present expansion
rate, and therefore we conclude that this effective age is similar
to the age of galaxies, or again of the order of $H_0^{-1}$ ($H_0$
being the present-day measured value of the Hubble ``constant'').
Therefore, the incompatibility argument in the core of the DT
argument is lost and reduces to the much weaker Fermi ``paradox'',
as we shall see in the further discussion.

Probably the more physical and meaningful way of restating the
entire situation is to reject the notion of an infinite age of the
Eddington-Lema\^{\i}tre model as a hollow formalism. A principle
sometimes ascribed to Aristotle or St. Augustine tells us that
there is no time without a changeable world. The state of perfect
equilibrium in the Eddington-Lema\^{\i}tre model in the $t \to
-\infty$ limit is exactly such an unchangeable state, without
means of determining either direction or the rate of passage of
time. In the sense of a modal version of the Aristotle-Augustine
principle, the temporal infinity in this model thus collapses into
a purely formal notion. Newton-Smith's formulation of this
principle:
\begin{quotation}
\noindent There is a period of time between the events E$_{1}$ and
E$_{2}$ if and only if relative to these events \textit{it is
possible} for some event or events to occur between
them,\footnote{Newton-Smith (1980), p. 44.}
\end{quotation}
\noindent explicitly points out to (macroscopic)
indistinguishability of moments in the state of complete
thermodynamical equilibrium. The same applies to the distant
future of the universe in which, according to many models, the
state of heat death is bound to occur. Barrow and Tipler (1978),
suggest that a formally infinite future should be substituted with
a finite interval, through an appropriate coordinate
transformation. A sort of counterexample, confirming the general
thesis that the cosmic time established by the Weyl postulate
should not be regarded as sacrosanct, is the diverging number of
(possible) events in the finite temporal vicinity of either the
initial or final global singularity. In such a situation a finite
cosmic time may be less appropriate than an alternative infinite
timescale (e.g.\ Misner 1969).\footnote{This obtained a poetic
description in Lord Byron's \textit{Cain\/} (1821):\par \par With
us acts are exempt from time, and we\par Can crowd eternity into
an hour,\par Or stretch an hour into eternity.} For instance, the
ever-decreasing number of events in the world approaching future
heat death (in the framework of some particular cosmological
model) could well be described, in the relationist picture, with
the finite time interval remaining; therefore, the time between
the initial singularity and the final heat death could be
represented by a (-$\infty $, 0) interval.

Is such a rescaling just a mathematico-philosophical perversion
lacking any relevance for the physical world? It seems that the
answer is firmly in the negative. While the elaboration lies
beyond the scope of the present paper, it is enough to point out
that the famous "biological scaling hypothesis" of Freeman Dyson
is just one guise of the re-scaling of time in the relationist
context (Dyson 1979). Many results in the nascent discipline of
physical eschatology depend on the Dyson's hypothesis, and it is
obvious that it's being true or else has a real, physical
consequences.

In brief, the past temporal infinity in the
Eddington-Lema\^{\i}tre model is \textbf{trivial} from the
anthropic point of view, and therefore DT argument is
inapplicable. Thus, one should reduce the realm of applicability
of the latter argument to cosmological models containing
\textbf{non-trivial past infinities}, i.e.\ an infinite chain of
non-trivial events.\footnote{Further discussion of the
non-triviality requirement can be found in Newton-Smith (1980). It
is important, for instance, in rejecting claims that statements of
the form ``the moment $t = t_{0}$ is now'' can \textbf{in
principle} describe any event whatsoever. Here we perceive the
connection between the anthropic argument and the relationist
\textit{vs.} absolutist controversy in the most plastic manner.}
The residual problem in each case is what is traditionally called
Fermi's (or the "Great Silence") paradox.

It should be immediately noted that DT argument, as outlined
above, is different from the unlimited entropy argument usually
used against cosmologies with past infinities (although the two
are related, as we shall see): why haven't irreversible processes,
in accordance with the thermodynamical laws, generated infinite
amounts of entropy in the universe by now? This is not just the
classical question of the thermodynamical disequilibrium between
the dark night sky and bright stars, but also the question of our
very existence, which is obviously contingent upon the large-scale
disequilibrium. Davies himself used the same argument against the
Hoyle-Narlikar conformally invariant cosmology\footnote{There is a
slight confusion in the literature as to which of several
different cosmological models is correctly called the
Hoyle-Narlikar cosmology. Here, we attach this name only to the
conformally invariant model with a conserved number of particles
and variable particle masses, such as exposed in Hoyle and
Narlikar (1972) and Hoyle (1975) papers. Although, as shown by
Narlikar and Arp (1993), several features of this model can be
incorporated into the revised steady state theory (which
postulates creation of matter, i.e. non-conservation of the
\textbf{number} of particles), we shall explicitly treat
Hoyle-Narlikar and the revised steady-state as separate theories,
since the distinction (conservation vs. creation) seems important
enough.} in his review of the latter in \textit{Nature} (Davies
1975), and Tipler (1982) mentions it in somewhat restricted sense,
as Olbers' paradox (again, expanded discussion may be found in
Barrow and Tipler 1986). The classical steady-state theory
alleviates this problem by the continuous creation of matter, and
an additional assumption that newly created matter is in a
low-entropy state. But cosmologies excluding creation of matter
(such as, for instance, the Einstein original static universe, or
the Hoyle-Narlikar conformally invariant cosmology) are faced with
this argument in a very serious form. Still, this thermodynamical
argument against steady-state models is qualitatively different
from DT argument we are dealing with, although both show how
difficulties arise when currently observable processes are
extrapolated backward in the past eternity. The latter argument is
based, essentially, on the diametrically opposed process: growth
of complexity, which results in emergence of technological
communities at some finite time (Kardashev and Strelnitskij 1988).
In the former case, we perceive increases in entropy in our
laboratory experiments; in the latter case, we perceive our
laboratories themselves as---in a sense---the very products of our
former observations.

In addition, attempts to reject the hypothesis of the infinite age
of the universe by a Kantian form of \textit{a priori} reasoning
are essentially different from the anthropic argument; the former
is also known as the \textit{kalam\/} cosmological argument (Craig
1979). One of the latest of these attempts has been made by Gerald
Whitrow (1978), which immediately caused many, chiefly negative,
reactions (Popper 1978; Bell 1979; Davies 1983; Gr\"{u}nbaum 1991;
Oppy 1995; see, however, Craig 1979, 1990). DT argument, on the
contrary, is not \textit{a priori}: it applies to the physically
well-defined sub-class of universes with infinite past and it is
firmly based in empirical (though non-standard) evidence---the one
on the existence of a technological civilization in vicinity of at
least one point of spacetime.

Thus, DT argument can be interpreted as the much stronger version
of the familiar ``Great Silence'' or ``astrosociological'' problem
(e.g.\ Brin 1983; Kardashev and Strelnitskij 1988; Almar 1989;
Gindilis and Rudnitskii 1993; Lipunov 1997; also known as the
Fermi's ``paradox''). Fermi's legendary question ``Where are
they?'' applies to the absence of any observable technologization
of the universe, as confronted by optimistic views on the
multitude of advanced extraterrestrial civilizations in our
Galaxy.\footnote{The similarities between the Olbers' and Fermi
``paradoxes'' has been discused by Almar (1989). For our purposes
the most important fact is that in both cases it is the finite age
of the universe and not it expansion which is the dominant
physical factor (Wesson et al. 1987).} As calculated by many
researchers (starting with the pioneering study of Hart 1975), the
timescale for the colonization of the Milky Way galaxy by a
technological society only a very little ahead of us in the
technological sense is very much smaller than the age of our
Galaxy. The age of the Galaxy is, of course, finite and became
rather well-known in recent years (Chaboyer et al. 1996, 1998;
Krauss 1998), and exactly that is the point at which the anthropic
argument is much more severe. As put dramatically by Lipunov
(1997):

\begin{quotation}
There are two observational... facts: (\ref{eq1}) the age of the
universe is $T = 10^{10}$ years and (\ref{eq2}) the time $\tau $
for the exponential development of our civilization is of the
order of some tens of years. For the sake of simplicity, we can
adopt $\tau = 100$ years, which is obviously an overestimate. A
gigantic dimensionless number arises, characterizing the growth of
a technological civilization over the time of existence of the
universe:
$$
K = \exp (T/\tau) \approx 10^{43,000,000}\; \; {\rm (!)}
$$
It is sufficient to say that theoretical physics has never dealt
with such large dimensionless numbers... In fact, it can be
confirmed that the probability of absence of ``space miracles'' in
our universe is $10^{-43,000,000}$, i.e., it is equal to zero!
Nevertheless, nobody has discovered them even after 20 years of
searches. On the contrary, a Great Silence of the universe has
been revealed.
\end{quotation}

The extension of this argument on both spatial and temporal scales
leads directly to the anthropic argument. Instead of a very large
age, we wish to investigate the limit $T \to \infty$ (in Lipunov's
presentation). Clearly, the DT argument (as well as practically
all other aspects of anthropic reasoning) is of great relevance
not only for cosmology, but for astrobiology and the prospects for
SETI, too. Since the latter topics are of potentially unprecedent
significance to the social and cultural history of human race,
this reaffirms the necessity of investigating all arguments of
relevance to the question of survival and evolutionary histories
of intelligent observers in the cosmological context. This is a
convenient point to reemphasize that (as noted among others, by
Lipunov himself) before Einstein and Friedmann---but after the
breakdown of the medieval creation dogma---there were no principal
differences between the two arguments. The traditional view of
XVIII, XIX and early XX century intellectuals has been that the
universe has always existed in conditions not very different from
those observable around us today. This makes the fact that the DT
argument appeared on the cosmological scene so late rather
strange.

\section{An Earth science interlude \label{hutt}}

Could it appear earlier? It is important to stress that the more
general problem of lack of information from the past has its
underground history deeply interwoven with the striving for
understanding natural history. The problem, as he have hinted, was
open as soon as the dogma of creation in 4004 BC was rejected,
openly or tacitly. It was, in fact, in the domain of Earth science
(rather than in astronomy) where the dam was broken: it is often
called the discovery of "deep time" or "geological time" (e.g.\
Ward 1998; Baxter 2004). And it was not accidental that the man
usually credited for this discovery, Scottish naturalist James
Hutton (1726$-$1797), was the first to reflect on the basic issue
of habitability of the indefinitely (or even infinitely) old
universe.

Following the lead of Gould (1987), we may approach Hutton's
solution of the problem of the duration of the past vs. limited
information transmitted from those epochs, noting how final causes
motivate the whole idea; today, we might cite the Strong Anthropic
Principle to the similar ends. Hutton imagined a "world machine":
his mechanicistic worldview found excellent field of applicability
in the geology of his day. Erosion of the soil is compensated by
the uplifting of mountains; any other particular tendency is
contrasted with an opposite one which is bound to return the world
to one or more previous stages. Hutton's vision is a geological
analog of the Empedoclean cyclic universe; hence the most famous
passage of his, ending the 1788 short version of his {\it Theory
of the Earth\/}:
\begin{quotation}
\noindent If the succession of worlds is established in the system
of nature, it is in vain to look for anything higher in the origin
of the earth. The result, therefore, of our present enquiry is
that we find no vestige of a beginning---no prospect of an
end.\footnote{Hutton (1788), p. 304.}
\end{quotation}
But, contrary to the standard textbook (and often quite Whiggish)
history, portraying Hutton as the standard-bearer of modern
scientific outlook, this view of the "world-machine" was not
motivated so much by the desire to explain the observed phenomena,
as by the metaphysical invocation of final causes. The final cause
in question was nothing less than what modern astrobiologists
would call \textbf{planetary habitability}: time and again, Hutton
writes of "mechanism of the globe, by which it is adapted to the
purpose of being a habitable world." The Earth was obviously
constructed (at some \textbf{indefinite} epoch, not existing from
eternity!) for higher purpose of being a habitat for life and,
eventually, for human domination. Hutton writes about
\begin{quotation}
\noindent ...a world contrived in consummate wisdom for the growth
and habitation of a great diversity of plants and animals; and a
world peculiarly adapted to the purpose of man, who inhabits all
its climates, who measures its extent, and determines its
productions at his pleasure.\footnote{Hutton (1788), pp.\
294-295.}
\end{quotation}
\noindent This seems absurd from the (Whiggish!) point of view of
modernity, but was almost self-evident in XVIII century; on the
other hand, Hutton's view bears striking similarity to some of the
modern teleological usages of the Strong Anthropic Principle we
shall discuss in \S\ref{tsap} below.

Note that the age of the Earth (and perhaps the rest of the
universe) was considered indefinite, but not infinite. Infinite
age would conflict with Hutton's profound Christian religiosity,
and he repeatedly implies that the ultimate questions of the
beginning and the end of the world are not part of the scientific
discourse. However, with profound subtlety, he builds an insurance
against Lucretian "eternal monuments" in his choice of words: not
that there is no beginning---there are only no \textbf{vestige} of
the beginning! The cyclic nature of the world machine erases the
relevant information from previous cycles and "cleans the slate".
But it is still internally inconsistent: if the world is made for
man, how comes that the achievements of previous generations of
humans are also erased? And if we accept pluralism about abodes of
life (which was rather standard in Hutton's time; see, for
instance, Fontenelle 1767), than it is very difficult to conceive
an explanation for the failure of intelligent beings to overcome
the slow processes of erosion and decay which erase information
from previous cycles. Empedocles at least postulated catastrophic,
singular events encompassing the entire universe; Hutton's world
machine is much less efficient in this respect. This dichotomy is
important to keep in mind before we return to the grand
cosmological scene.

\section{Classical Steady-State Theory}

In order to better assess the importance of DT argument and its
application to the cosmologies with infinite past, we shall
briefly consider the conceptual foundations of the most famous and
historically most influential such cosmology, the 1948. model of
Hermann Bondi, Thomas Gold and Fred Hoyle (Bondi and Gold 1948;
Hoyle 1948). Although there is some controversy whether the
classical steady-state cosmology represented a single entity or
two disjointed theories (that of Bondi-Gold and the version of
Hoyle), we shall refer to them as the classical steady-state
model, discussing, where relevant, particular differences among
the two versions (Hoyle 1949). While Hoyle's version is generally
superior, being formulated in the language of the classical field
theory, for our purposes it is, in fact, the perfect cosmological
principle (henceforth PCP) of Bondi and Gold (1948) that makes the
important point most clearly. Its essentially non-mathematical
character makes it even more transparent in the sense of giving
the core formulation of uniformitarianism in cosmology (Balashov
1994).

One fact that remained largely overlooked is that the classical
steady-state cosmology displayed one of the very first instances
of anthropic reasoning in modern science. In the founding paper,
Bondi and Gold (1948) gave a specific anthropic flavor to the
classical unlimited entropy argument ("Olbers' paradox"):

\begin{quotation}
A static universe would clearly reach thermodynamical equilibrium
after some time. An infinitely old universe would certainly be in
this state. There would be complete equilibrium between matter and
radiation, and (apart possibly from some slight variations due to
gravitational potentials) everything would be at one and the same
temperature. There would be no evolution, no distinguishing
features, no recognizable direction of time. That our universe is
not of this type is clear not only from astronomical observations
but from local physics and indeed \textit{from our very
existence}. [present author's emphasis]
\end{quotation}

This way of reasoning is not only deeply founded in the normative
physical practice, but is directly responsible for the conceptual
simplicity of the classical steady-state, praised even by its
adversaries such as Sir Martin Ryle. However, it is also important
to perceive that the paragraph quoted above contains a
characteristic example of the anti-Empedoclean double standard
deeply rooted in modern science. Namely, our existence is taken
into account in physical theory when it is convenient, in this
particular example when it comes to proving that the entropy of
the universe is far from the maximal value. At the same time,
other consequences of our existence as a technological
civilization, which lead to the DT argument, are conveniently
ignored. It can be hardly contested that it was exactly our
capability to adapt and technologize nature which led, among all
other things, to advances in mathematics and astronomy leading to
the formulation of the classical steady-state theory. In what
follows, we shall show how the DT argument is operationalized in
the context of this theory.

PCP formulated by Bondi and Gold (1948), can be simply expressed
as the homogeneity of the universe in 4-dimensional spacetime.
This is just the generalization of so-called Cosmological
principle (cf.\ Milne 1940), which assumes homogeneity in space,
but not necessarily in time. Mathematically speaking, PCP can be
formulated as a necessity to have a timelike Killing vector in the
classical Robertson-Walker metric. Thus, PCP leads to the
line-element of the well-known (de Sitter) form (in usual $c = 1$
units):
\begin{equation}
\label{eq1}
ds^2 = dt^2 - e^{2Ht}\left[ {\,dr^2 + r^2\left( {\,d\vartheta ^2 + \sin
^2\vartheta \,d\varphi ^2} \right)} \right],
\end{equation}
\noindent where $H$ is the true constant, and can take any real
value. Now, $H = 0$ leads to a static universe, which can be
discarded not only on clear observational grounds, but (even more
interesting from our point of view) from the thermodynamical
considerations as well. Olbers' paradox testifies that the
universe has not reached the state of thermodynamical equilibrium,
which is impossible to avoid in an infinitely old static
cosmological model (see the celebrated classic discussion in Bondi
1961).\footnote{The $H = 0$ case corresponds to infinite euclidean
static universe, similar to the Einstein original static model
(Einstein 1917). The difference lies in topology, since the
Einstein model is topologically closed. However, the Einstein
model also (albeit trivially) satisfies the Perfect Cosmological
Principle. Although we can not delve deeper into this topic, it is
worth noticing that DT argument applies to flat euclidean as well
as Einstein closed static models. The absence of the large-scale
expansion in these static models makes the expansion of life and
intelligence significantly easier. Even the hierarchical
distribution of galaxies and supergalactic structures, like in
Charlier original fractal model (Charlier 1922; Kalitzin 1961) or
in Segal's chronometric cosmology (e.g.\ Segal 1978), does not
alleviate this problem. No distance is out of reach in an
infinitely old static universe. Moreover, we may speculate,
following Barrow and Tipler (1986), that the absence of global
relativistic effects like shear and torsion in such universes
makes the technologization of ever larger spatial volumes even
more important, since there are no negentropy sources for
information processing other than the matter fields. Of course,
the basic problem of all static cosmologies is Olbers' paradox,
i.e.\ global thermodynamical disequilibrium, so in a sense the
existence of even a single intelligent observer is
\textit{reductio ad absurdum\/} of such cosmologies!} The case $H
< 0$, which corresponds to the universal contraction, presents a
situation in which the radiation of distant sources is shifted to
the violet end of the spectrum, resulting in an infinitely bright
sky background in the manner still less acceptable than in the
case of Olbers' paradox in the static universe. Therefore, the
only possible conclusion is that $H > 0$, which is the realistic
case of the expanding universe.

Part of the appeal of the steady-state concept can be found in
words of Sciama (quoted according to Kragh 1996):
\begin{quotation}
\noindent The steady-state theory opens up the exciting
possibility that the laws of physics may indeed determine the
contents of the universe through the requirement that all features
of the universe be self-propagating... The requirement of
self-propagation is thus a powerful new principle with whose aid
we see for the first time the possibility of answering the
question why things are as they are without merely saying: it is
because they were as they were.
\end{quotation}
However, as we shall see below, the germ of doom lies exactly in
the concept of self-propagation, since it seems to be incapable of
correctly accounting for a specific "feature" of the universe,
namely us. In other words, if we accept the Empedoclean picture
(in which the biological and mental evolution is an inherent and
necessary part of the cosmological evolution), than although it
has to be self-propagating, the rise of intelligence at the same
time \textbf{must not} be self-propagating.

The basic violations of uniformity we empirically notice in the
universe are galaxies. Newly created matter is
\textbf{continuously }condensed in galaxies, and although the
details of this process have remained controversial (Sciama 1955;
Harwit 1961), mainly because insufficient theoretical work was
devoted to it prior to the universal rejection of the steady-state
picture in mid-1960s, the predictive power of PCP is manifested
here once again. The answer offered by the steady-state outlook to
the question of the age distribution of galaxies on a sufficiently
large scale is essentially independent of physical details of
galaxy formation. In the classical steady-state model, the
distribution function of galaxies is simply
\begin{equation}
\label{eq2} f\left( x \right) = e^{-3Ht},
\end{equation}
where H is the Hubble constant, a true constant in
contradistinction to the Friedmann models. Taking into account Eq.
(\ref{eq2}) the average age of galaxies is simply
\begin{equation}
\label{eq3}
\left\langle \tau \right\rangle = \frac{1}{n}\int\limits_0^\infty {3nH\,e^{
- 3H\alpha }d\alpha } {\kern 1pt} \, = \frac{1}{3}H^{ - 1}.
\end{equation}
This illustrates a beautiful simplicity which PCP imposes on the
theory: the average age of galaxies is calculated without any
reference to the complicated physics of galaxy
formation.\footnote{Parenthetically, modern Big Bang cosmologies
are still uncertain as to the average \textbf{predicted} age of
galaxies within a factor of about two (see, for instance, Gott
1977; Peebles 1993). Note that this is something very different
from the \textbf{observational} uncertainties in determination of,
say, age of the oldest globular clusters in the Galaxy. The latter
is in practice regarded as \textit{limes inferior\/} of the age of
the universe. The two are sometimes confused, especially in
popular scientific literature, due to the hegemonic position of
the standard hot Big Bang paradigm. \textit{A priori}, it is not
necessary that age of Galaxy (and galaxies) is determined by
cosmological factors at all; classical steady-state cosmology is a
good such counterexample. In it the age of any chosen galaxy is a
random variable. However, in the standard model it is the
irreversibility of the Hubble expansion (i.e.\ the cosmological
``arrow of time'') which necessarily links cosmology with galactic
cosmogony. This, of course, does not imply that the classical
steady-state model does not have inherent difficulties with the
physics of galaxy formation (e.g.\ Harwit 1961).} However, we
should keep in mind a historical fact of great importance: the
estimates of the Hubble constant relevant in the late 1940s were
in gross violation of what we today know as the plausible interval
for that quantity. For $H \sim  500$ km s$^{-1}$, which was the
then reigning Hubble measurement, the Friedman-Lema\^{\i}tre
cosmology (which soon became, following a derogatory comment of
Fred Hoyle in a BBC radio broadcast, the Big Bang cosmology) has
had a serious conflict with the age of the Earth and chemical
elements (e.g.\ Bok 1946). At the time of formulation of the
steady-state theory, $\langle \tau \rangle $ was considered small
($\sim 6 \times 10^8$ yrs, due to the gross overestimate of the
value of Hubble constant), and the Milky Way has already been an
extraordinary old galaxy, which certainly implied that surrounding
galaxies are far less probable to achieve the same degree of
chemical and biological evolution. Although this circumstance in
fact does not alleviate the DT argument, it certainly does have a
significant psychological effect, making problems with
technologization literally much more distant.

The fraction $\delta (t)$ of galaxies older than the age $t$ is
given by
\begin{equation}
\label{eq4} \delta \left( t \right)=
\frac{1}{n}\int\limits_t^\infty {3nH\,e^{ - 3Hx}{\kern 1pt} dx =
e^{-3Ht},}
\end{equation}
This is the mathematical root of the problem, reflecting the fact
that the exponential function is everywhere finite. For instance,
if we take $\langle \tau \rangle $ to be an order of magnitude
higher, in accordance with the today's best knowledge on the
magnitude of the Hubble constant, DT argument quoted above gains
force. Since the fraction of galaxies that are older than age $t =
2 \times  t_{MW} \approx 2 \times 1.2 \times  10^{10}$ yrs is
given by Eq.\ (\ref{eq4}), it follows that there are almost 2{\%}
of all galaxies in any large enough comoving volume which are
\textbf{twice} the age of the Milky Way. We should keep in mind at
all times that our Galaxy is \textbf{already} old enough for the
Fermi's paradox to be formulated (as briefly discussed in \S 2).

The issue of different predictions of ages of galaxies in
different cosmological models has been present on the cosmological
scene since the very beginning of controversy between the
classical steady-state and what will ultimately be called the
standard Big Bang model. This problem is attractive because of the
possibility of observational verification on spatial scales
smaller than those required for most of the other cosmological
tests. Studies of convenient age indicators in a large enough
sample of galaxies could, in principle, be performed in order to
answer the question whether galaxies have approximately the same
age, or span a wide distribution, as given by Eq.\ (\ref{eq2}). In
practical terms, however, the task is extremely difficult, since
even the age of the Milky Way (by far the best known galaxy, of
course) has not been known with less than 20{\%} uncertainty until
very recently. Therefore this sort of empirical evidence has not
been actually used very much against the classical steady-state
model. The age-based attack on the steady-state theory was
initiated by Gamow, who in a short note in 1954 (Gamow 1954),
pointed out that after the Baade's revision of cosmological
distance and timescales, steady-state theory faces the problem of
underestimating ages of galaxies in our vicinity. However, it is
clear in the Gamow paper that he does not find the so far measured
galactic ages very convincing. The same age discrepancy issue has
been raised by the great American observational astronomer, Ivan
King, who in 1961 pointed out that the ages of stellar populations
in most of nearby galaxies are estimated to be about $2H_0^{-1}$,
which is not in agreement with the first order prediction of the
steady-state theory (King 1961). In a short, but very
comprehensive reply to this objection, Hoyle and Narlikar have
suggested several weaknesses in this argument and put forward at
least one problem which outlived the cosmological controversy and
remains puzzling to this day (Hoyle and Narlikar 1962). First of
all, they pointed out that judging confidently the age of galaxies
means knowing with certainty the evolutionary effects dictated by
intragalactic physics. As a crucial example, they offer uncertain
status of the morphological types of galaxies: are they constants
for all times, or do they change with cosmic time? The latter
alternative precludes any conclusion based on assumptions relating
stellar populations with the morphological type (a standard
procedure in astrophysics) unless we know the exact law of
evolutionary change. It is interesting to note that Hoyle and
Narlikar in connection with this point suggest a scheme of
transformation of Hubble's morphological types which is still an
acceptable hypothesis today. Some observational indications to
that effect, as well as theoretical explanations bearing on the
nature of dark matter, have only recently been reported (Braine
and Combes 1993; Pfenniger, Combes and Martinet 1994).

Finally, Hoyle and Narlikar in their 1962 paper indicate (in close
connection with their subsequent ``radical departure''; see Hoyle
and Narlikar 1966) the possibility of temporal correlations due to
collective effects. This idea will be realized in detail in the
so-called revised or quasi-steady state model of 1990s. If
galaxies are formed in groups (i.e.\ on the higher level of
structure), it may be assumed that the galactic ages are
correlated. Therefore, it is natural to expect that galaxies in
the vicinity of the Milky Way (those amenable to detailed
observations and age measurements) will have similar ages (both
among themselves, and in comparison with our Galaxy).

\section{Closed Steady State Models}

As we have discussed in the introductory part of this study, the
DT argument has been used for the first time against a stationary
cosmological model with closed topology advanced by George F. R.
Ellis and his collaborators (Ellis 1978; Ellis et al.\ 1978). A
similar cosmological model has been developed in the mid-1990s by
American physicist Peter Phillips (1994a, b). These models,
sharing several key similarities, we shall call the \textbf{closed
steady-state models}. Their basic characteristic is that
stationarity is achieved by rejection of the ``usual''
cosmological principle (not to mention PCP). In other words, the
temporal steady-state is paid for by abandoning spatial
homogeneity.\footnote{In this sense, they do not contradict the
statement that the classical steady-state model is unique, i.e.
that there is only one cosmology satisfying PCP, as Bondi was fond
of emphasizing (e.g.\ Bondi 1961). This methodological advantage
of the classical steady state model over all other cosmological
models remains unscathed. As we shall see, closed steady state
models are forced to invoke WAP for explaining the highly special
nature of our view of the universe. For the same task, in both the
Friedmann models and in the classical steady-state case, Occam's
razor is strong enough.} It is a matter of philosophical taste
whether one considers the price too high or not.\footnote{This is
quite independent from the fact that closed steady state
cosmologies do not satisfy observational constraints and can be
considered rejected today. As far as the plausibility of the
hypothesis that the observable universe is a homogeneous part of
the much larger inhomogeneous whole, see for instance Harwit
(1995). This feature is inherent to some variants of the
inflationary scenario, most notably Linde's chaotic inflation
program.} However, it is difficult to avoid being disturbed by
elaboration of Ellis et al.\ that
\begin{quotation}
\noindent while isotropy is directly observable, homogeneity (on a
cosmological scale) is not. In the standard discussions the
assumption of homogeneity is made \textit{a priori}, either
directly, or in some equivalent form (e.g.\ as the assumption that
the Universe is isotropic for \textit{all\/} observers...), and so
is not subjected to observational verification. Accordingly, the
standard 'proof' of the expansion of the Universe is based on an
unverified \textit{a priori\/} assumption.
\end{quotation}

With the intention of investigating consequences of abandoning the
homogeneity postulate and retaining Einstein field equations,
Ellis and his collaborators have reached a model of a
topologically closed universe with two privileged ``points''. We
are located near one them, and that is not accidental, because (in
accordance with WAP) it is expected for us to be located in the
regions possessing the necessary properties for the origination
and evolution of complex (biological) systems. This occurs near
the ``center'' of the universe (it is most natural to use the term
for our pole of the manifold, by analogy with a 3-sphere).
Opposite of the center is located the singularity surrounded by
hot matter, simulating in this manner the initial singularity in
the Friedmann models. However, in this static model, the
singularity is \textbf{co-present} with everything that exists,
not preceding it. Obviously, it makes the model more appealing
from the epistemological point of view: although laws of nature
break down at singularity, in Ellis et al. model it is not forever
inaccessible in the past, but could, in principle, be investigated
using the methods and apparatus of modern science. This co-present
singularity can be intuitively understood as being the
``enclosure'' or ``mantle'' surrounding the universe. Its major
purpose is to play the role of a recycling facility in the global
cosmological ecology, since the static nature of the universe
makes recycling of high-entropy matter necessary. In the framework
of this model it is achieved through a streaming of high-entropy
matter (mainly in the form of heavy elements synthesized in
stellar nucleosynthesis) toward the singularity, where it is
dissociated and returned to the universe in the form of
low-entropy matter (presumably hot single baryons). In this manner
the total entropy stays the same at all epochs. Beside this
streaming (which does not change the net mass distribution), there
is no systematic motion: observable redshift is of purely
gravitational origin. The co-present singularity in Ellis et al.\
model bears a resemblance to the \textit{apeiron\/} of
Anaximandros out of which worlds are formed and unto which they
ultimately dissociate.

In the cosmological model of Phillips (1994a, b) there are also
two singular points, this time called the northern and southern
pole. The Milky Way galaxy is located in close proximity to the
northern pole of the universe (from the same anthropic reasons as
in the Ellis et al. model). In contradistinction to the Ellis et
al. model, here we have a systematic motion of galaxies, in
direction from the northern to the southern pole. This motion,
however, is laminar and stationary, so that the universe in
general always offers the same picture to a typical observer.
Metric coefficients are independent of time, and in this sense the
model could be considered static. This is a situation somewhat
similar to the famous de Sitter cosmological solution for an empty
universe, which is nominally static, although we now interpret it
as describing the exponential expansion. Since this large-scale
motion is present, the observed redshift is partially of
gravitational and partially of Doppler origin. In the Phillips'
model there are two postulated types of matter, which he calls
primary and secondary matter, where the matter to which we are
accustomed to is of secondary type. Primary matter is moving in
the opposite direction (from the southern to the northern pole),
so that the generalized form of matter conservation is preserved,
while the two types of matter never interact except in singular
points at the poles. The details of thermodynamics of this model
have been elaborated in better detail than in the Ellis et al.\
model (Phillips 1994b), wherein some observational tests of this
version of closed stationary models have been proposed.

These empirical tests are fatal to closed stationary models. In
the very original paper of Ellis et al.\ (1978), it is shown that
this model is not able to account properly for the so-called ($m,
z)$ curve, i.e.\ the relationship between the apparent magnitude
and redshift of cosmologically distributed sources of radiation.
It is harder to disprove Phillips' model, since the gravitational
and Doppler redshifts are delicately entangled. The cleanest test
could be the measurement of peculiar motion of distant sources
with respect to the universal reference frame as defined by the
microwave background radiation. This experiment is possible to
perform in the case of rich galaxy clusters, by means of the
Sunyaev-Zeldovich effect (Sunyaev and Zeldovich 1980). The
prediction of Phillips' model is that more distant clusters will
tend to have significantly larger peculiar motions than those
nearby. Recent measurements indicate that this is not the case,
and there is no meaningful way to save the theory (Phillips 2001,
private communication). Therefore, we may consider stationary
closed cosmologies to be rejected by observations.

However, they are useful for us in the historical sense. DT
argument against them continues to hold and is even more forceful
than for the case of the classical steady-state. It is reasonable
to assume that in closed universes, the cross-section for contact
(and technologization) of an advanced civilization could literally
cover a large fraction of the entire universe. Since our position
is necessarily privileged in these universes (hence we observe
galaxies and the limiting singularity to be nearly isotropic
around our position), it is only plausible to assume that the same
anthropic reasons which establish such a situation are acting for
\textbf{any} intelligent community which could \textbf{ever} arise
in such a universe. But in that case, where the choice of places
of birth of intelligent observers is necessarily finite, it is
very easy to see that the world line of any civilization older
than ours will pass through the present of the Solar system. And
in these models, again, conditions for the emergence of
intelligent observers (in a limited spatial region, admittedly)
persist for an infinite time, and arbitrarily old civilizations
are a possibility. The difference in the nature of large-scale
motions in the two models considered is irrelevant for our
purposes, since the peculiar motions of nearby (with respect to
any intelligent observers) galaxies are in any case much smaller
than the rate by which the contact cross-section for advanced
civilization increases.

One could imagine the situation in which an advanced civilization
emerges in a privileged region of space (that is, in the vicinity
of our present position), and gradually expands to encompass the
entire ``favorable'' spatial region (being of finite size, as just
a part of the finite universe). It is reasonable to argue that in
the infinite past such scenario happened at least once [by the
same token as the conclusions of Ellis and Brundrit (1979) apply
with respect to spatial infinity]. There are two possible
follow-ups. Such supercivilization could either exist for a
definite, or for indefinite period of cosmic time. In the first
alternative, the universe we currently inhabit must be
``recycled'', as Davies warned; no mechanism for such a
``de-technologization'' is known or even envisaged at present.
Since we can envisage (if only very vaguely) the methods through
which advanced communities of intelligent beings may technologize
ever larger spatial volumes of the universe, and in the case of a
topologically closed universe, even the entire such universe (e.g.
Tipler 1994), a humble Humean approach suggests that we choose a
``smaller miracle''---that a supercivilization can exist for an
indefinite time.

In any case, we may safely conclude that in the universes of
finite size and infinite age (as modelled by Ellis et al. and
Phillips' theories) the anthropic argument necessarily leads us to
paradoxes, if only we do not restrict the growth of complexity,
socio-technological advance of intelligent societies and their
technologization of the environment by definitional \textit{fiat}.
It goes without saying that the situation is equally grave for the
other model universes in which the universe is infinite in both
spatial and temporal extension, but in which the conditions
favorable to life persist in either a finite or infinite region
for an infinite time. The main lesson of the anthropic spatial
selection such as our proximity to the boreal pole in Phillips'
model is that this form of self-selection allows for most of the
universe to be uninhabitable, and still retain DT argument. This
is valid even in the case in which the universe is infinite and
uninhabitable except for the finite region around our present
location.

As we can see in retrospect, the very fact of applying DT argument
against closed steady-state theories demonstrates that PCP is too
strong a requirement for the operation of the argument. As Tipler
(1982) stressed, only stationarity and limited local properties
are required. From a philosophical point of view, it should be
noted that it is also necessary that the rise of intelligent
communities and their expansion are \textbf{possible} within a
given astrophysical environment. This is self-evident, since the
humanity exists for a finite time in a relatively stable
environment, and expansion over interstellar or even intergalactic
lengthscales is, if not yet a reality, at least quite conceivable
from our point of view. The ultimate reason for this is our
empirical knowledge on the constancy of physical laws and their
modes of operation over these lengthscales. In a strongly
inhomogeneous universe, or universe with random fluctuations on
the scales of, say, 1 pc, the argument looses its power. However,
according to the WAP selection, while free to ask questions about
possible physical origin of such a hypothetical bizarre behaviour,
we should not seek to confirm our conjectures by performing
experiments and observations in the real world, because such
specific circumstance would preclude our existence (Earman 1987).
Therefore, it seems that in the cases similar to the Ellis et al.
and Phillips' cosmologies, the DT argument can not justifiably be
regarded as contingent on anything stronger than WAP, as Barrow
and Tipler (1986) tend to do.

\section{The Teleological SAP ``Counterargument'' \label{tsap}}

One of the possible recourses for a steady-state proponent in this
quandary concerns invoking the Strong Anthropic Principle
(henceforth SAP) in its teleological interpretation. It is
necessary, therefore, to pause for a moment and consider the
meaning and possible interpretations of the SAP, since it has been
and still is a considerable source of confusion in the field of
anthropic research. In the famous exposition of Carter (1974),
several important anthropic principles were defined. Among them,
the most speculative and thought-provoking has been exactly the
SAP which states that:
\begin{quotation}
\noindent ...the Universe (and hence the fundamental parameters on
which it depends) must be such as to admit the creation of
observers within it at some stage.
\end{quotation}

Unfortunately, there is no such thing as \textbf{the} definition of SAP.
Widely used definition of Barrow and Tipler (1986) in their influential
monograph, has somewhat different overtones from the Carter's original:

\begin{quotation}
\noindent The Universe must have those properties which allow life
to develop within it at some stage in its history.
\end{quotation}

As noted and discussed in some detail by Earman (1987) and Bostrom
(2002), these definitions are not equivalent, and the one in
Barrow and Tipler certainly possesses (as, parenthetically, the
entire monograph) strong teleological overtones. While one can
plausibly argue that there is in fact nothing particularly
``strong'' about SAP in the Carter's formulation if it is regarded
as pointing to \textbf{blanks} for future physical explanation
(Balashov 1990), the somewhat heated discussion of this issue is
outside of the scope of the present study. Without entering into a
general debate on the merits and shortcomings of the teleological
discourse in cosmology, it should be noted that it arose as a
natural reaction to the overuse of the Copernican principle in
natural sciences, and particularly in cosmology. From this overuse
follows, for instance, uncritical (and often even unconscious)
acceptance of cosmological homogeneity, discussed in the section
devoted to closed steady state models. From a prejudice that
nothing in our position is special one may draw bizarre
conclusions, for example that it requires explanation that we are
not right now located in the intergalactic space, since the latter
fills more than 99.99{\%} of the volume of the universe, and any
spatial location not in it is truly exceptional on a grand scale.
Understanding that we live in \textit{a priori} very improbable
universe\footnote{One should, for instance, keep in mind the
estimate of Penrose that \textit{a priori} probability of Big Bang
happening in so smooth (low gravitational entropy) manner as to
produce the observable universe is only 1 part in
$10^{10^{123}}$!} is an encouragement to teleological projects of
various kinds, of which not all must be unscientific (Barrow and
Tipler 1986; Tipler 1994). One of strategies for refuting the
usage of the DT argument lies exactly in assuming that the
appearance of intelligent observers is not only of low
probability, but in the literal sense impossible. This has been
acknowledged, among others, by Ellis and Brundit (1979), who
concluded that ``the existence of life on our own planet does
\textit{not\/}  prove that this probability is non-zero''. With
that kind of approach, our existence is a miracle, which has
happened for some inexplicable, theological reason. This is an
extreme anti-Empedoclean attitude in the framework of which the
biological (or at least anthropological and psychological)
evolution is completely transcendent compared to the physical one.
The ontological gap between the two seems irreducible. This
attitude is traditionally (although in a shallow, and sometimes
openly incorrect, interpretation) linked to the major religious
doctrines, but it is interesting to note that the same sort of
thinking is to be found in writings of thinkers of opposite---or
at least anticlerical---orientation, such as Sir Fred Hoyle.

In his extraordinarily interesting and well-written
autobiographical reminiscences, Hoyle (1994) writes, in connection
with his anthropic prediction of the $^{12}$C level, but also,
probably, alluding to some of the stranger consequences of his own
steady-state outlook:\footnote{A popular account of this problem
and the Hoyle's answer may also be found in Davies' book on
physical eschatology (Davies 1994).}

\begin{quotation}
\noindent All of this suggested to me what I suppose might be
called profound questions. Was the existence of life a result of a
set of freakish coincidences in nuclear physics? Could it be that
the laws of physics are not the strictly invariant mathematical
forms we take them to be? Could there be variations in the forms,
with the Universe being a far more complex structure than we take
it to be in all our cosmological theories? If so, life would
perforce exist only where the nuclear adjustments happened to be
favorable, removing the need for arbitrary coincidences, just as
one finds in the modern formulation of the weak anthropic
principle. Or is the Universe teleological, with the laws
deliberately designed to permit the existence of life, the common
religious position? A further possibility, suggested by the modern
strong anthropic principle, did not occur to me in 1953 -- namely,
that it is our existence that forces the nuclear details to be the
way they are, which is essentially the common religious position
taken backwards. Before ridiculing this last possibility, as quite
a few scientists tend to do, it is necessary, as I pointed out
before, to explain the condensation of the universal wave function
through the intervention of human consciousness.
\end{quotation}

Even more explicit is the discussion presented in his
philosophically-oriented review article written in 1982:

\begin{quotation}
\noindent In {\it Steady State Cosmology Revisited\/} (University
College Cardiff Press, 1980) I estimated (on a very conservative
basis) the chance of random shuffling of amino acids producing a
workable set of enzymes to be less than $10^{-40000}$... Rather
than accept a probability less than 1 in $10^{40000}$ of life
having arisen through the ``blind'' forces of nature, it seems
better to suppose that the origin of life was a deliberate
intellectual act. By ``better'' I mean less likely to be wrong...

Suppose you were a superintellect working through possibilities in
polymer chemistry. Would you not be astonished that polymers based
on the carbon atom turned out in your calculations to have the
remarkable properties of the enzymes and other biomolecules? Would
you not be bowled over in surprise to find that a living cell was
a feasible construct? Would you not say to yourself, in whatever
language supercalculating intellects use, "Some supercalculating
intellect must have designed the properties of the carbon atom,
otherwise the chance of my finding such an atom through the blind
forces of nature would be less than 1 part in $10^{40000}$." Of
course you would, and if you were a sensible superintellect you
would conclude that the carbon atom is a fix.
\end{quotation}

In this manner, to the Adam's dilemma (from \textit{The Paradise
Lost}), one answers in affirmative: yes, the celestial bodies
truly exist for the sake of the Earth and human beings. Hutton
would be happy with this solution! The explanation of absence of
extraterrestrial technology on large scales (that is, both Fermi's
paradox in standard cosmology and DT problem in an eternal
universe) lies in the fact that the probability of the spontaneous
conception and subsequent evolution of any intelligent observers
is \textbf{exactly zero}.\footnote{It seems that here one may find
a fault in Hoyle's reasoning. He estimates, on the biochemical
bass, that the probability of a spontaneous assembly of the first
living cell is smaller than $10^{-40000}$, and from this draws a
conclusion that the universe must be much older than $H_0^{-1}$,
as well as that a variation of the classical panspermia hypothesis
must be a correct explanation for the presence of life on Earth.
However, the panspermia hypothesis only increases the available
4-volume for random physical processes to bring about life; this
volume still remains finite. On the other hand, the universe of
the classical steady-state theory (as well as Hoyle's later
revised steady-state alternative) is truly infinite in both space
and time.} The very title of important article of Kardashev and
Strelnitskij (1988), \textit{Supercivilizations as possible
products of the progressive evolution of matter\/} is simply
\textbf{wrong} on this view, since there are underlying
\textbf{physical} reason for impossibility of transition between
evolution of matter and that of life, and later mind itself
(presumably leading to the state of ``supercivilization''). Before
we reject this as an obsolete and ridiculously dogmatic viewpoint,
one should note that this view is probably the only way known so
far capable of \textit{prima facie\/} accounting for the often
overlooked Wigner's result that, within the quantum mechanical
formalism, the probability of spontaneous creation of living
systems is equal to zero (Wigner 1967, p.\ 200). Quantum
mechanical considerations also motivated Hoyle, in particular in
his 1982 paper.

As pointed out by Tipler (2001, private communication), this
picture is difficult to defend along several lines. Major problem
with this sort of argument is that it assumes that not merely
intelligent life, but the specific species \textit{Homo sapiens},
is doing the selecting of the actual universe. There is no
positive reason for belief that humanity is ultimately privileged
in this way for participation in ``creation'' of the universe. On
the contrary, as the father of the very label ``anthropic
principle'' Brandon Carter emphasized, every form of intelligent
life is in exactly the same situation. Of course, the purpose of
such Carter's opinion is exactly to excise, together with
anthropocentrism, the teleological mode of explanation, and the
answer we consider in this section is manifestly teleological.
However, it is not necessary to be dogmatic either way in
considering these issues. This danger may be avoided on account of
a general attitude (see, for instance, the discussion in Sklar
1985) that a metaphysical conjecture can be accepted as an
explanatory hypothesis, if capable of accounting for the existing
empirical evidence, and in particular in cases where the empirical
evidence is slim.

Along these guidelines, one may speculate that more intelligent
and more advanced species would be better at ``creating''
universes than we could ever hope to be (Harrison 1995). Even in
the cases of manifestly teleological schemes, such as the
Omega-point theory of Tipler (1994), there is nothing inherently
advantageous in belonging \textbf{specifically} to \textit{Homo
sapiens sapiens}. Our future descendants, the beings who will
ultimately realize the purpose of the universe in reaching the
Omega point, can not with certainty be characterized as closer to
us than we are to birds and fishes. By openly recognizing the
issue of the melioristic universe in all its ramifications, Tipler
(1994) shows that teleology needs not necessarily to be burdened
by dogmas of the times past. The same lesson should, undoubtedly,
apply to the teleological mode of accounting for DT argument in
universes with an infinite past series of events.

Fortunately, we are \textbf{not} in the actual position to choose
between these alternatives, since our observable universe is
certainly of finite age. However, this dilemma can resurface if we
find that the larger whole in which the observable universe is
just an embedded part, possesses a structure characterized by the
past temporal infinity. We shall return to this point in \S 7. For
the moment, it should be noted that there is another possible
recourse for the steady-state picture, which Hoyle has used in the
latest phase of his cosmological thinking. This is the argument of
the quasi- or revised steady-state theory developed by him,
Burbidge and Narlikar in a series of papers published in 1990s
(Hoyle 1992; Hoyle, Burbidge, and Narlikar 1993, 1994). Here we
basically have a novel strategy in the fight with entropy, and we
now investigate whether it is any more successful in dealing with
DT argument.

\section{Quasi-Stationarity and Intelligent Species}

Although CSS theory is now universally considered defunct, there
are some recent developments to be considered in light of the
preceding analysis. Open of them is the emergence of the "Quasi"
steady-state theory (henceforth QSS) as an attempt to overcome the
difficulties with observational evidence against an unchangeable
universe of PCP, and in favor of a hot state in the cosmic past
(Hoyle et al. 1993). The roots of this attempt can be found in
Hoyle's early work (Hoyle 1949) in which, comparing the merits of
his and Bondi-Gold version of CSS, he writes:

\begin{quotation}
\noindent Bondi and Gold... in discussing the continuous creation
of matter, have avoided the introduction of a quantitative theory
by making the hypothesis that the universe, when taken on a
sufficiently large scale satisfies the wide (perfect) cosmological
principle. Since the wide cosmological principle is very far from
being satisfied over regions with linear dimension less than about
10$^{24}$ cm, such a hypothesis immediately raises the following
questions: What considerations determine the scale necessary for
the wide cosmological principle to be approximately satisfied?
\end{quotation}

Insofar as we can detect inhomogeneities in the universe in the
form of galaxies, clusters and other large-scale structure, we can
set the spatial scale for the application of the (restricted)
cosmological principle. It is not \textbf{obvious} which intervals
similarly characterize smoothing of temporal fluctuations. Of
course, the relevant scale can be introduced by a definitional
\textit{fiat}, thereby effectively introducing a new constant of
nature, which is generally unsatisfactory. Taking amplitudes of
such fluctuations to be similar to the conventional Hubble time
directly led Hoyle (with fascinating consequence) to the idea of
mini-big bangs in the revised steady-state theory. Without going
into technical details (presented in Hoyle and Burbidge 1992;
Hoyle et al. 1993, 1994), it should be mentioned that, as well as
in Hoyle's version of CSS model, negative energy of the creation
field transforms into matter with positive energy. However, the
creation is not uniform in spacetime, but occurs in discrete
creation events, so-called ``mini-bangs''. In each individual
``mini-bang'' about $10^{16}$ Solar masses (a characteristic mass
of superclusters of galaxies) is created in the form of particles
with Planck mass ($M_{Pl} \sim 10^{-5}$ g). Each Planck particle
ultimately produces about $5 \times 10^{18}$ baryons which react
at high energies producing light chemical elements. The
distribution of creation events creates the characteristic
cellular structure seen in the large galaxy surveys of the last
decade. The thermal energy of matter expanding from the creation
events is the ultimate origin of the all-pervading cosmic
microwave background; its anisotropy and local departure from the
thermodynamical equilibrium are lost through repeated interactions
with specific form of cosmic dust: elongated metallic whiskers
(``needles'') created and distributed through space by the
supernovae explosions of the first generation of stars.

The revised steady state theory possesses a continuity with the
earlier work of Hoyle and Narlikar, mainly their results from
1966. study (Hoyle and Narlikar 1966), where the rejection of
smooth continuous creation of the previous theory has been
emphasized in the very title of the paper (\textit{A radical
departure from the 'steady-state' concept in cosmology}). This
model, in its elaboration of 1990s, includes some of the elements
of recent---particularly observational
astrophysical---developments, and so is more modern and closer to
the prevailing trends in science. At the same time, however, its
structure is complicated and possesses none of the beautiful
simplicity of the classical steady-state theory. Although detailed
discriminatory observational tests of the new model have not been
performed yet, the probability that they will give results
expected by the QSS proponents is small indeed. Moreover, the
findings connected with the chemical abundances of the primordial
matter, as well as the deep galaxy fields, point to the lack of
support for some of the basic tenets of this model.

However, QSS does not stand much better against the anthropic
arguments. This theory is vulnerable to the DT argument in the
same basic manner as CSS cosmology. The fact that we shall
encounter galaxies of similar age to the Milky Way inside of the
Local Supercluster represents only a gigantic spatial and temporal
translation of the problem, which does not brings us closer to its
solution. Galaxies and technological civilizations of the
appropriate age will be present in other superclusters, which
could be of an arbitrary age, obeying only the self-similar
"supercluster distribution function" necessarily akin to Eq.\
(\ref{eq2}). This strategy of ``passing the buck'' is probably one
of the chief reasons why DT argument has needed so much time to be
formulated. It may eventually solve our psychological difficulty
concerning past temporal infinity, but not the physical problem
itself. If we do not forbid information transport between the
superclusters by definitional \textit{fiat}, the problem remains
as acute as in the classical steady-state case.

\section{Stationary Multiverse?}

Recently the idea of stationarity on very large scales has been
reanimated in the form of several similar inflationary scenarios.
A typical example is the work of Andrei Linde and his
collaborators (Linde 1988, 1990; Linde et al. 1994), as well as
Vilenkin (1992, 1995). In these models (known under labels such as
``chaotic'' or ``eternal'' inflation), bubbles are formed out of
spacetime foam at Planck energy, each bubble evolving into an
individual universe in its own right, with specific topologies,
geometries, laws of nature, coupling constants, etc. The entire
process of separation and inflation of these individual
bubble-universes has no beginning or end, and therefore the entire
ontological background of these processes (for which the
appropriate name of \textbf{multiverse} is coined) is stationary.
A significant similarity between classical C-field cosmology and
inflationary scenarios in general has been noted recently (Hoyle
1992; Narlikar 1984). The manifestation of that swing of the
pendulum backward from extreme evolutionism toward some form of
stationarity can be seen in the very titles of several recent
papers, such as \textit{From the Big Bang Theory to the Theory of
Stationary Universe }(Linde et al. 1994). \footnote{This
phenomenon in its historical context could be compared to the
similar return of (neo) catastrophism on the scene in geology and
paleontology (e.g.\ Clube 1995).} It is still too early to
estimate whether this should be regarded as the general tendency
to recover some of the advantages of "stationary" cosmologies
(and, possibly, a counter-reaction to the overemphasis on
"evolutionary uncertainties" frequently employed in astrophysics
as an \textit{ad hoc}, or rather \textit{ignoramus}, recipe).
Future historians of science will have to discuss this question.

The role of ``local inhomogeneities'' which is played by galaxies
in the classical steady-state model, is played by entire
individual bubble-universes in the multiverse theories. As
suggested by Linde (1990), different ways of breaking of the
initial complete symmetry of the single ``unified'' force of
nature will occur in different bubble-universes, and so a wild
variety of physical conditions are likely to arise. It is natural
to ask, therefore, whether the DT anthropic argument applies to
those quantum cosmological models which are in a global stationary
state. It should be immediately clear that the inflationary
scenarios have great relevance for the entire problem of existence
of life in the universe. As an illustration, in a conclusion to
their highly technical paper on some aspects of inflation, Novello
and Heintzmann (1984) write:

\begin{quotation}
\noindent Two possibilities arise to have a sufficiently old
universe: either $S_0$ [the present-day scale factor of our
universe] is large---which means that the Universe was never very
dense and thereby never very hot (this would guarantee biological
conditions for all of the cosmic epoch); or $1/H_0$ is very large
and $S_0$ small. In this case biological reactions will only occur
(or reoccur) in the late expanding phase and existence of life
would only occur at a finite time; whereas in the first case, life
could have existed eternally in the universe, leading to the
intriguing hypothesis that there may be colonies in space which
are infinitely more intelligent than we are.
\end{quotation}

There are several methods to escape the conclusions of the DT
argument in the multiverse case. The simplest one is to reject the
very notion of ``eternal'' inflation. This possibility
(differently motivated) has been investigated by Vilenkin (1995)
and Borde and Vilenkin (1994). The goal of these studies is to
show that, under a general range of physical conditions, the
initial singular beginning is compulsory, from which it
immediately stems that the entire multiverse is of finite age. The
outcome is not unambiguous. Another possible approach is the one
discussed by Barrow and Tipler in Chapter IX of their monograph:
to forbid information transport between individual bubble-universe
by physical reasons. The ``environment'' surrounding the
bubble-universes is false vacuum at energies close to the Planck
energy, which makes any communication hard to imagine, to say the
least (but see Garriga et al.\ 1999). This conclusion may be
challenged on the grounds that (i) what is hard to imagine today,
does not need to be so for the advanced civilizations which have
already technologized a large part or their entire domicile
bubble-universe and thus marshaled unbelievable material and
intellectual resources, and (ii) it may well be possible to
communicate between two bubble-universes in a non-classical way,
that is using (or creating!) the complex topological structure of
the multiverse. The latter method would rely on some version of
the well-known concept of ``wormholes'' (e.g.\ Morris, Thorne and
Yurtsever 1988; Visser 1990). Finally, the third possible strategy
lies in possibility that individual bubble-universes (with finite
resources, and therefore the finite duration of an active
technologized state) may be separated from the Planck spacetime
foam at a rate sufficiently high enough to achieve a state
equilibrium with the formation and evolution of generic
intelligent communities. Such an equilibrium will reflect the
state in which most of ``young'' universes are uninhabited either
by their native supercivilizations or by an external
(``colonizing'') supercivilization of an arbitrary age. However,
it goes without saying that this is extremely speculative, and
therefore is mentioned here just for the sake of completness.
Otherwise, one cannot escape an agreement with the judgement of
Barrow and Tipler that
\begin{quotation}
\noindent this... objection is much weaker in the inflation
steady-state universe situation than it is in the standard
steady-state universe model, for it is far from clear that it is
possible to develop technology which will allow intelligent life
to exist... in the steady state region.
\end{quotation}

\section{Lessons and morals}

From the mind-boggling reaches of the multiverse inflationary
theories, let us return to the classical cosmology and try to
summarize our conclusions so far. Several classical cosmological
models have been compared in Table 1 with respect to some
properties relevant for the anthropic reasoning and survival of
intelligent observers. These properties refer to both dynamics,
topology and application of the DT argument. Taken together, they
illustrate the range (certainly not exhaustive) of classical
cosmological thought and the close connection of the entire
anthropic reasoning to other physically defined properties of
particular models. All models presented in Table 1 possess a past
temporal infinity in at least one sense (i.e.\ according to one of
the major theories of the ontological status of time discussed
above). Some of them are static at large scales, and most are not;
the ambiguous sign for the case of classical de Sitter universe
stands for the curious historical ambiguity on the meaning of
``static''.\footnote{Metric coefficients do not change with time
in this model, and for that reason it was conceived and received
as static, at least in the first decades after 1917; according to
the modern view, it is an exponentially expanding model.} Some
models are topologically closed, while others are flat or open,
and some possess such specific features as (global) singularities
and horizons.\footnote{It is important to emphasize that the
singularities under considerations are the global ones; otherwise,
all spacetimes in which at least some worldlines are incomplete
should qualify as singular, including the classical steady state
theory which is characterized by creation ``micro-singularities''
(and does not manifestly prevent formation of black hole
singularities!).}

\begin{table}[htbp]
\begin{tabular}
{|p{15pt}|p{60pt}|p{30pt}|p{33pt}|p{40pt}|p{44pt}|p{43pt}|p{28pt}|}
\hline \textbf{No.}& \textbf{Model}& \textbf{Static}&
\textbf{Closed}& \textbf{Matter}& \textbf{Singular}&
\textbf{Horizon}&
\textbf{D-T} \\
\hline 1. & Einstein & +& +& +& -& -&
+ \\
\hline 2. & de Sitter& + -& -& -& -& +&
- \\
\hline 3. & Eddington-Lema\^{\i}tre& -& -& +& -& +&
- + \\
\hline 4. & Classical SS& -& -& +& -& +&
+ \\
\hline 5. & Ellis et al.& +& +& +& +& -&
+ \\
\hline 6. & Phillips& -& +& +& +& -&
+ \\
\hline 7. & Oscillatory nonsingular & -& +& +& -& -&
+ - \\
\hline 8. & Revised SS & -& -& +& -& +&
+ \\
\hline
\end{tabular}
\label{tab1} \caption{Comparison of some classical cosmological
models on several counts related to the anthropic argument against
past temporal infinities.}
\end{table}

In the last column of Table 1, the applicability of the DT
argument, according to our study, is presented. In the case of the
Eddington-Lema\^{\i}tre model, the outcome strongly depends on the
detailed physical properties of the initial equilibrium state. In
most versions of this model, the anthropic argument is clearly
inapplicable, although the strong version of the argument may be
operational in some specific versions. The dependence on the
physical detail also plays a crucial role in considerations of
non-singular cyclical models (singular models can not be
justifiably called oscillatory at all). Here the real issue is
information transport between the successive phases of contraction
and expansion. If the physical conditions are too extreme at the
point of the smallest radius, the information on the previous
cycle will be erased, and the subsequent phase will essentially be
a completely new universe, ``from scratch''. These conditions, in
turn, depend on the exact values of cosmological parameters in the
model,\footnote{And if these parameters are allowed to vary
between the cycles in a random manner, there is no reason to
believe in an infinite number of cycles at all: the cosmological
constant, which is one of these parameters, could change the sign
and become strongly positive, which would lead to an
ever-expanding (although topologically still closed!) universe.}
as well as in unknown physics on which the concept of a bounce
critically depends. In addition, the duration of expanding and
contracting phase must be long enough for the advanced
technological communities to arise (WAP constraint), or---in the
stronger version---for the process of spontaneous creation of
probes (and probes causally connected with the region beyond the
bounce at that!) to occur. Since various combinations of
parameters are possible in this class of models, it is not
possible to give a generic answer on the applicability question.

A criticism usually encountered from people for the first time
facing such an anthropic argument concerns the relationship of
laws and instances in cosmology. How can such an argument based on
a single instance of intelligent and technological life be used as
a general argument against a wide spectrum of specific physical
cosmologies, such as the classical steady-state theory certainly
is. There are two possible answers to this basic \textit{non
sequitur}. First of all, there is no generic form of cosmological
model; as shown clearly by Balashov (1994) in the detailed
analysis of CSS theory, the entire theory \textbf{can} be both
legitimately and practically derived from an essentially
methodological principle as is PCP. \textit{Mutatis mutandis}, the
rejection of such a theory can be both legitimately and
practically done on the account of argument pointing out the
self-contradictory nature of particular instances of PCP
application. Let us reiterate: there is no inherent need for life
to exist in the universe satisfying the Perfect Cosmological
Principle. In a sense, it would be much easier to formulate PCP
for the counterfactual case of lifeless universe, since amplitudes
of ``local inhomogeneities`` could as well be either much smaller
or much larger from the narrow range required for satisfying the
WAP constraints. For instance, the counterfactual universe with
amplitude of fluctuations several orders of magnitude smaller than
those detected in the real universe through cosmic microwave
background observations could evolve as mildly inhomogeneous
plasma according to PCP without ever forming galaxies and stars.
New matter will simply keep the density of such plasma constant.
Very high (again counterfactual) value of the Hubble constant will
probably have the same effect. However, once we make an
observation of existence of life (and intelligence), PCP requires
that we have the same in all ages and in infinitely many places.

The second aspect of the dilemma is contained in the ambiguous
status of the biological and psychological evolution in our
picture of the universe. It seems clear that we are dealing with a
double standard here. When Fred Hoyle predicted the existence of
7.65 MeV metastable level in the nucleus of $^{12}$C on the basis
of existence of carbon and carbon-based life (Hoyle et al. 1953;
Hoyle 1994), it was quite clear that this level does not exist
separately from the entire scope of physical sciences. On the
contrary, it was clear from the beginning that the energy and all
relevant properties of this level can \textbf{in principle} be
reduced to the particular numerical values of the constants of
nature, like the Planck constant and elementary charge. Therefore,
fine tuning of the $^{12}$C level is actually only a
\textbf{manifestation} of the fine tuning of various constants of
nature. Conceptually, this manifestation is redundant. However, we
can not, as discussed in Barrow and Tipler (1986), back track and
reconstruct the manner in which the particular values of constants
propagate toward creating the observed energy of this level. This
is just a technical matter, because the carbon nucleus is an
extremely complicated quantum system, and there is simply not
enough sophistication and calculating power in the present-day
nuclear physics to perform such back tracking. Still, it is
important to note that nobody denies \textbf{principal}
possibility of such reconstruction, which can be achieved in a
matter of decades if the development of numerical science and
computers continues at the present pace. The same idea applies,
\textit{mutatis mutandis}, to other examples of so frequently
discussed ``anthropic coincidences''.

However, a significant resistance is encountered when the same
reasoning applies to systems which are more complex than
nuclei---say living and intelligent beings. Still, there is no
evidence whatsoever that living systems are in any way different
from usual physical systems except in the level of complexity
(Davies 1999). Views sometimes expressed to the contrary can be
regarded only as remnants of the obsolete vitalistic doctrines.
Indications that the mental processes occurring in intelligent
beings can be \textbf{in principle } ultimately reduced in a
manner essentially the same as is conjecture for the carbon
nucleus have been gathered by Tipler (1994) and Stapp (1985).
Similar thoughts have been expressed earlier in XX century by
Erwin Schr\"{o}dinger in his influential \textit{What is Life? }
essays (Schr\"{o}dinger 1944).

It is clear that the weak version of DT argument may be criticized
on historico-sociological arguments. Let us assume that there is
an absolute maximum for the development of any community of
intelligent beings, and that the contact cross section of this
maximum is small. In that case, if the rate of emergence of
intelligent communities is sufficiently low in comparison to the
Hubble expansion rate, in the finite relaxation time it is
possible to achieve the equilibrium in which arbitrarily large
fraction of the comoving 3-volume is non-technologized. This could
be called the Spenglerian model of intelligent communities (e.g.
Fischer 1989). For an exception to the prevailing paradigm of
discontinuity between the contemporary worldview in the natural
sciences and history understood in Spenglerian terms, one may look
at the excellent recent study of Victor Clube devoted to cometary
neocatastrophism in planetary studies (Clube 1995). This work,
under the instructive title \textit{The Nature of Punctual Crises
and the Spenglerian Model of Civilization} presents---in a
different, still quite relevant scientific framework---how much
the self-contentedness and dogmatical blindness in science in the
last century or two could lead toward a \textit{de facto} wrong
factual road. It was first the medieval immutable heavens, and
subsequently uniformitarian (anticlerical, ironically enough)
dogma on exclusiveness of slow evolutionary change, which
impeded---and to a degree still impedes---acceptance of the truth
about the significance of catastrophic events originating in our
cosmic environment. The specific problem with the Spenglerian
model lies exactly in the mentioned need for
\textbf{non-exclusivity}: as the teleological counter-argument
becomes valid only if the probability of spontaneous creation of
life (and/or von Neumann probes) is for some reason exactly equal
to zero, and not only very small, so here the number of
civilizations which are capable of escaping the Spenglerian
``curse'' must be exactly zero in order for the explanation to
work in an infinitely old universe. This is even harder to imagine
than the previous case, because if the proponents of impossibility
of spontaneous creation of life can enlist quantum mechanics on
their side, the proponents of the Spenglerian model have to show
that sociology and history are in a sense more universal than
quantum mechanics itself! (Of course, the Spenglerian model
remains a strong palliative to naive concepts of universe
blossoming with life in case of temporally and spatially limited
systems; therefore, it remains very relevant for the Fermi's
paradox and the SETI problem in the Milky Way.) Of course, the
stronger Tipler's version of the anthropic argument, related to
spontaneously assembled von Neumann probes is entirely immune to
this type of criticism.

The core lesson of the entire case for the anthropic argument
against cosmologies containing past temporal infinities is,
however, located on a deep epistemological level. As a side effect
of both the Copernican revolution and the Cartesian dualism, the
implicit rejection of the pre-Socratic picture of the
inseparability of the cosmological, biological and anthropological
domains led to an inevitable delay in noticing a powerful and
specific cosmological argument. Further discussions on this topic,
as well as further discussions of the future of \textbf{physical
universe}, will have to \textbf{explicitly} take into account the
existence and activities of intelligent observers. This will
manifest itself not only in retrodictions about the cosmological
past, as the original anthropic argument of Dicke and Carter has
been traditionally used, but also through the predictive aspect of
cosmology. These considerations will necessarily be of a
multidisciplinary character, so desirable in this latest epoch of
development of our picture of the universe.

\vspace{0.4cm}

\section*{References}
\refe Almar, I. 1989, \textit{Acta Astronautica\/} \textbf{26},
253.

\refe Balashov, Yu. 1990, \textit{Comments Astrophys.}
\textbf{15}, 19.

\refe Balashov, Yu. 1994, \textit{Studies in History and
Philosophy of Science} \textbf{25B}, 933.

\refe Barrow, J. D. and Tipler, F. J. 1986, \textit{The Anthropic
Cosmological Principle} (Oxford University Press, New York).

\refe Baxter, S. 2004 {\it Ages in Chaos : James Hutton and the
Discovery of Deep Time\/} (Forge Books, New York).

\refe Bekenstein, J. D. and Meisels, A. 1980, \textit{Astrophys.
J.} \textbf{237}, 242.

\refe Bell, J. 1979, \textit{Brit. J. Phil. Sci. }\textbf{30},
161.

\refe Bok, B. J. 1946, \textit{Mon. Not. R. astr. Soc.
}\textbf{106}, 61.

\refe Bondi, H. 1961, \textit{Cosmology} (2$^{nd}$ edition,
Cambridge University Press, London).

\refe Bondi, H. and Gold, T. 1948, \textit{Mon. Not. R. astr.
Soc.} \textbf{108}, 252.

\refe Borde, A. and Vilenkin, A. 1994, \textit{Phys. Rev. Lett.}
\textbf{72}, 3305.

\refe Bostrom, N. 2002, {\it Anthropic Bias: Observation Selection
Effects\/} (Routledge, New York).

\refe Braine, J. and Combes, F. 1993, \textit{Astron. Astrophys.}
\textbf{269}, 7.

\refe Brin, G. D. 1983, \textit{Q. Jl. R. astr. Soc.} \textbf{24},
283.

\refe Burnet, J. 1908, \textit{Early Greek Philosophy} (Adam and
Charles Black, London).

\refe Carter, B. 1974, in \textit{Physical Cosmology and
Philosophy}, ed. by Leslie, J. 1990. (Macmillan, London).

\refe Chaboyer, B., Demarque, P., Kernan, P. J., Krauss, L. M. and
Sarajedini, A. 1996, \textit{Mon. Not. R. astr. Soc.
}\textbf{283}, 683.

\refe Charlier, C. V. L. 1922, \textit{Ark. Mat. Astron. Phys.
}\textbf{16}, No. 22.

\refe Clube, S. V. M. 1995, \textit{Vistas in Astronomy}
\textbf{39}, 673.

\refe Collins, C. B. 1990, \textit{Class. Quantum Grav.}
\textbf{7}, 1983.

\refe Cornford, F. 1965, \textit{Thucydides Mythistoricus\/}
(Greenwood Press Publishers, New York).

\refe Craig, W. L. 1979, \textit{The Kalam Cosmological Argument
}(Macmillan, London).

\refe Craig, W. L. 1990, \textit{Brit. J. Phil. Sci. }\textbf{41},
473.

\refe \'Cirkovi\'c, M. M. 2000, \textit{Serb. Astron. J.}
\textbf{161}, 33.

\refe \'Cirkovi\'c, M. M. 2003a, \textit{Found. Phys.}
\textbf{33}, 467.

\refe \'Cirkovi\'c, M. M. 2003b, \textit{Astron. Astrophys. Trans}
\textbf{22}, 879.

\refe Davies, P. C. W. 1975, \textit{Nature} \textbf{255}, 191.

\refe Davies, P. C. W. 1978, \textit{Nature} \textbf{273}, 336.

\refe Davies, P. C. W. 1983, \textit{God and the New Physics}
(Simon {\&} Schuster, New York).

\refe Davies, P. C. W. 1994, \textit{The Last Three Minutes}
(Basic Books, New York).

\refe Davies, P. C. W. 1999, \textit{The Fifth Miracle} (Simon
{\&} Schuster, New York).

\refe Dicke, R. H., Peebles, P. J. E., Roll, P. G. and Wilkinson,
D. T. 1965, \textit{Astrophys.} $J.$ \textbf{142}, 414.

\refe Diels, H. 1983, \textit{Presocratic Fragments} (Naprijed,
Zagreb).

\refe Dyson, F. J. 1979, {\it Rev. Mod. Phys.} {\bf 51}, 447.

\refe Earman, J. 1987, \textit{Am. Phil. Quart. }\textbf{24}, 307.

\refe Eddington, A. S. 1928, \textit{The Nature of the Physical
World} (Cambridge University Press, London).

\refe Eddington, A. S. 1930, \textit{Mon. Not. R. astr. Soc.
}\textbf{90}, 668.

\refe Einstein, A. 1917, \textit{Sitz. Preuss. Akad. Wiss., } 112,
142.

\refe Ellis, G. F. R. 1978, \textit{Gen. Rel. Grav.} \textbf{9},
87.

\refe Ellis, G. F. R., Maartens, R. and Nel, S. D. 1978,
\textit{Mon. Not. R. astr. Soc.} \textbf{184}, 439.

\refe Ellis, G. F. R. and Brundrit, G. B. 1979, \textit{Q. Jl. R.
astr. Soc.} \textbf{20}, 37.

\refe Fairbanks, A. 1898, \textit{The First Philosophers of
Greece\/}(K. Paul, Trench {\&} Trubner, London).

\refe Fischer, K. P. 1989, \textit{History and Prophecy - Oswald
Spengler and the Decline of the West} (Peter Lang, New York).

\refe Fontenelle, Bernard Le Bouyier de 1767, {\it Conversation on
the Plurality of the Worlds\/} (transl. from French, 2nd edition,
London).

\refe Gamow, G. 1954, \textit{Astrophys. J.} \textbf{59}, 200.

\refe Garriga, J., Mukhanov, V. F., Olum, K. D. and Vilenkin, A.
2000, {\it Int. J. Theor. Phys.} {\bf 39}, 1887.

\refe Gindilis, L. M. and Rudnitskii, G. M. 1993, in \textit{Third
Decennial US-USSR Conference } \textit{on SETI}, ed. by Seth
Shostak, G. (ASP Conference Series, Vol. 47, San Francisco), p.
403.

\refe Gott, J. R. 1977, \textit{Ann. Rev. Astron. Astrophys.}
\textbf{15}, 235.

\refe Gould, S. J. 1987, {\it Time's Arrow, Time's Cycle\/}
(Harvard University Press, Cambridge).

\refe Gr\"{u}nbaum, A. 1991, \textit{Erkenntnis}, \textbf{35},
233.

\refe Guthrie, W. K. C. 1969, \textit{A History of Greek
Philosophy II\/} (Cambridge University Press, London).

\refe Haldane, J. B. S. 1945, {\it Nature\/} {\bf 155}, 133.

\refe Harrison, E. R. 1973, \textit{Ann. Rev. Astron. Astrophys.
}\textbf{11}, 155.

\refe Harrison, E. R. 1995, \textit{Q. J. R. astr. Soc.
}\textbf{36}, 193.

\refe Hart, M. H. 1975, \textit{Q. J. R. astr. Soc. }\textbf{16},
128.

\refe Harwit, M. 1961, \textit{Mon. Not. R. astr. Soc.}
\textbf{122}, 47.

\refe Harwit, M. 1995, \textit{Astrophys. J. }\textbf{447}, 482.

\refe Hawking, S. W. and Penrose, R. 1970, \textit{Proc. Roy. Soc.
A} \textbf{314}, 529.

\refe Hawkins, D. 1971, \textit{Philosophy of Science}
\textbf{38}, 273.

\refe Hinckfuss, I. 1975, \textit{The Existence of Space and Time
}(Clarendon Press, Oxford).

\refe Hoyle, F. 1948, \textit{Mon. Not. R. astr. Soc.}
\textbf{108}, 372.

\refe Hoyle, F. 1949, \textit{Mon. Not. R. astr. Soc.}
\textbf{109}, 365.

\refe Hoyle, F. 1975, \textit{Astrophys. J.} \textbf{196}, 661.

\refe Hoyle, F. 1982, \textit{Ann. Rev. Astron. Astrophys.
}\textbf{20}, 1.

\refe Hoyle, F. 1992, \textit{Astrophys. and Space Sci.}
\textbf{198}, 195.

\refe Hoyle, F. 1994, \textit{Home Is Where the Wind Blows}
(University Science Books, Mill Valley).

\refe Hoyle, F., Dunbar, D. N. F., Wensel, W. A. and Whaling, W.
1953, \textit{Phys. Rev. }\textbf{92}, 649.

\refe Hoyle, F. and Narlikar, J. V. 1962, \textit{Observatory}
\textbf{82}, 13.

\refe Hoyle, F. and Narlikar, J. V. 1966, \textit{Proc. Roy. Soc.
A} \textbf{290}, 162.

\refe Hoyle, F. and Narlikar, J. V. 1972, \textit{Mon. Not. R.
astr. Soc.} \textbf{155}, 305.

\refe Hoyle, F. and Burbidge, G. R. 1992, \textit{Astrophys. J.
}\textbf{399}, L9.

\refe Hoyle, F, Burbidge, G. R. and Narlikar, J. V. 1993,
\textit{Astrophys. J.} \textbf{410}, 437.

\refe Hoyle, F, Burbidge, G. R. and Narlikar, J. V. 1994,
\textit{Mon. Not. R. astr. Soc.} \textbf{267}, 1007.

\refe Hutton, J. 1788, {\it Theory of the Earth. Transactions of
the Royal Society of Edinburgh\/} {\bf 1}, 209.

\refe Israelit, M. and Rosen, N. 1989, \textit{Astrophys. J.
}\textbf{342}, 627.

\refe Kalitzin, N. S. 1961, \textit{Mon. Not. R. astr.
Soc.}\textbf{ 122}, 41.

\refe Kardashev, N. S. and Strelnitskij, V. S. 1988, in
\textit{Bioastronomy -- The Next Steps}, ed. by Marx, G. (Kluwer,
Dordrecht), p. 295.

\refe King, I. 1961, \textit{Observatory} \textbf{81}, 128.

\refe Kragh, H. 1996, \textit{Cosmology and Controversy}
(Princeton University Press, Princeton).

\refe Krauss, L. M. 1998, \textit{Astrophys. J. }\textbf{501},
461.

\refe Linde, A. D. 1988, \textit{Phys. Lett. B} \textbf{211}, 29.

\refe Linde, A. D. 1990, \textit{Inflation and Quantum Cosmology}
(Academic Press, San Diego).

\refe Linde, A. D., Linde, D. and Mezhlumian, A. 1994,
\textit{Phys. Rev. D} \textbf{49}, 1783.

\refe Lipunov, V. M. 1997, \textit{Astrophys. and Space Sci.}
\textbf{252}, 73.

\refe Lucretius 1997, \textit{On the Nature of Things} (translated
by William E. Leonard, e-text version, Project Gutenberg, Urbana).

\refe Milne, E. A. 1948, \textit{Kinematic Relativity} (Clarendon
Press, Oxford).

\refe Misner, C. W. 1969, \textit{Phys. Rev.} \textbf{186}, 1328.

\refe Morris, M. S., Thorne, K. S. and Yurtsever, U. 1988,
\textit{Phys. Rev. Lett.} \textbf{61}, 1446.

\refe Narlikar, J. V. 1984, \textit{J. Astrophys. Astr.
}\textbf{5}, 67.

\refe Narlikar, J. V. and Arp, H. 1993, \textit{Astrophys. J.}
\textbf{405}, 51.

\refe Newton-Smith, W. H. 1980, \textit{The Structure of Time}
(Routledge and Kegan Paul, London).

\refe North, J. 1965, \textit{The Measure of the Universe: A
History of Modern Cosmology} (Oxford University Press, London).

\refe Novello, M. and Heintzmann, H. 1984, \textit{Gen. Rel.
Grav.} \textbf{16}, 535.

\refe O'Brien, D. 1969, \textit{Empedocles' Cosmic Cycle}
(Cambridge University Press, Cambridge).

\refe Oppy, G. 1995, \textit{Faith and Philosophy} \textbf{12},
237.

\refe Peebles, P. J. E. 1993, \textit{Principles of Physical
Cosmology} (Princeton University Press, Princeton).

\refe Pegg, D. T. 1971, \textit{Mon. Not. R. astr. Soc.
}\textbf{154}, 321.

\refe Penrose, R. 1979, in \textit{General Relativity: An Einstein
Centenary}, ed. by Hawking, S. W. and Israel, W. (Cambridge
University Press, Cambridge), p. 581.

\refe Pfenniger, D., Combes, F. and Martinet, L. 1994,
\textit{Astron. Astrophys.} \textbf{285}, 79.

\refe Phillips, P. R. 1994a, \textit{Mon. Not. R. astr. Soc.}
\textbf{269}, 771.

\refe Phillips, P. R. 1994b, \textit{Mon. Not. R. astr. Soc.}
\textbf{271}, 499.

\refe Popper, K. R. 1978, \textit{Brit. J. Phil. Sci.
}\textbf{29}, 47.

\refe Price, H. 1996, \textit{Time's Arrow and Archimedes' Point}
(Oxford University Press, Oxford).

\refe Reber, G. 1982, \textit{Publ. Astr. Soc. Australia},
\textbf{4}, 482.

\refe Schr\"{o}dinger, E. 1944, \textit{What is Life? }(Cambridge
University Press, Cambridge).

\refe Sciama, D. W. 1955, \textit{Mon. Not. R. astr. Soc.
}\textbf{115}, 3.

\refe Segal, I. E. 1978, \textit{Astron. Astrophys.} \textbf{68},
343.

\refe Sklar, L. 1985, \textit{Philosophy and Spacetime Physics}
(University of California Press, Berkeley).

\refe Smolin, L. 1992 {\it Class. Quantum Grav.} {\bf 9}, 173.

\refe Spengler, O. 1918, \textit{Decline of the West }(1996
edition by Alfred A. Knopf Publisher, New York).

\refe Stapp, H. P. 1985, \textit{Found. Phys.} \textbf{15}, 35.

\refe Sunyaev, R. A. and Zeldovich, Ya. V. 1980, \textit{Ann. Rev.
Astron. Astrophys.} \textbf{18}, 537.

\refe Tipler, F. J. 1981, \textit{Q. Jl. R. astr. Soc.}
\textbf{22}, 133.

\refe Tipler, F. J. 1982, \textit{Observatory} \textbf{102}, 36.

\refe Tipler, F. J. 1994, \textit{The Physics of Immortality}
(Doubleday, New York).

\refe Vilenkin, A. 1992, \textit{Phys. Rev.} $D$ \textbf{46},
2355.

\refe Vilenkin, A. 1995, \textit{Phys. Rev. Lett.} \textbf{74},
846.

\refe Visser, M. 1990, \textit{Phys. Rev. D }\textbf{41}, 1116.

\refe Ward, P. D. 1998, {\it Time Machines: Scientific
Explorations in Deep Time\/} (Copernicus Books, New York).

\refe Weinberg, S. 1972, \textit{Gravitation and Cosmology}
(Wiley, New York).

\refe Wesson, P. S., Valle, K. and Stabell, R. 1987,
\textit{Astrophys. J.} \textbf{317}, 601.

\refe Whitrow, G. J. 1978, \textit{Brit. J. Phil. Sci.
}\textbf{29}, 39.

\refe Wigner, E. 1967, \textit{Symmetries and Reflections}
(University of Indiana Press, Bloomington).
\end{document}